%% file: SUS-19-004_temp.tex
\pdfoutput=1
\documentclass[11pt,twoside,a4paper,cmspaper,final,collab]{cms-tdr}
\def\svnVersion{86f1ffe-D}\def\svnDate{2021/07/19}\def\cmsCernNoTag{CERN-EP-2021-015}\def\cmsCernDate{\today}\def\cmsMessage{"Published in Physical Review D as \href{http://dx.doi.org/10.1103/PhysRevD.104.032006}{\doi{10.1103/PhysRevD.104.032006}.}"}
\begin{document}\cmsNoteHeader{SUS-19-004}

\newlength\cmsTabSkip\setlength{\cmsTabSkip}{1ex}
\ifthenelse{\boolean{cms@external}}{\providecommand{\cmsLeft}{upper\xspace}}{\providecommand{\cmsLeft}{left\xspace}}
\ifthenelse{\boolean{cms@external}}{\providecommand{\cmsRight}{lower\xspace}}{\providecommand{\cmsRight}{right\xspace}}
\ifthenelse{\boolean{cms@external}}{\providecommand{\NA}{\ensuremath{\cdots}\xspace}}{\providecommand{\NA}{\ensuremath{\text{---}}\xspace}}
\ifthenelse{\boolean{cms@external}}{\providecommand{\CL}{C.L.\xspace}}{\providecommand{\CL}{CL\xspace}}

\newcommand{\njets}{\ensuremath{N_\text{jets}}\xspace}
\newcommand{\snn}  {\ensuremath{S_\mathrm{NN}}\xspace}
\newcommand{\snni}[1]  {\ensuremath{S_\mathrm{NN,#1}}\xspace}
\newcommand{\sgo}{\ensuremath{\widetilde{\mathrm{S}}}\xspace}
\newcommand{\lsp}{\PSGczDo}
\newcommand{\singlet}{\HepParticle{S}{}{}\xspace}
\newcommand{\mstop}{\ensuremath{m_{\PSQt}}\xspace}
\newcommand{\syy}  {\ensuremath{\singlet \HepParticle{Y}{}{} \HepAntiParticle{Y}{}{} }\xspace}

\cmsNoteHeader{SUS-19-004} 
\title{Search for top squarks in final states with two top quarks and several light-flavor jets in proton-proton collisions at \texorpdfstring{$\sqrt{s}=13\TeV$}{sqrt(s) = 13 TeV}}

\date{\today}

\abstract{
  Many new physics models, including versions of supersymmetry characterized by 
  $R$-parity violation (RPV),
  compressed mass spectra, long decay chains, or additional hidden sectors, predict the production of 
  events with top quarks, low missing transverse momentum, and many additional quarks or gluons. 
  The results of a search for new physics in events with 
  two top quarks and additional jets are reported. 
  The search is performed using events with at least seven jets and exactly one electron or muon. 
  No requirement on missing transverse momentum is imposed. 
  The study is based on a sample of proton-proton collisions at 
  $\sqrt{s} = 13\TeV$ corresponding to 137\fbinv of integrated luminosity collected with the 
  CMS detector at the LHC in 2016--2018.  
  The data are used to determine best fit values and upper limits 
  on the cross section for pair production of top squarks in 
  scenarios of RPV and stealth supersymmetry.
  Top squark masses up to 670 (870)\GeV are excluded at 95\% confidence level for the 
  RPV (stealth) scenario, and the maximum observed local signal significance is 2.8 standard 
  deviations for the RPV scenario with top squark mass of 400\GeV.    
}

\hypersetup{
pdfauthor={CMS Collaboration},
pdftitle={Search for top squarks in final states with two top quarks and several light-flavor jets in proton-proton collisions at sqrt(s) = 13 TeV},
pdfsubject={CMS},
pdfkeywords={CMS,  SUSY}}

\maketitle 

\section{Introduction}
\label{sec:intro}

Supersymmetry~\cite{susy, Martin:1997ns} (SUSY) is an extension of the standard model (SM) 
that may provide a solution to the gauge hierarchy problem~\cite{Dimopoulos:1995mi}.
In the SUSY framework, quadratically divergent radiative corrections to the Higgs boson 
mass parameter, dominated by loops involving the top quark, are canceled by loops with
bosonic top quark superpartners (top squark, \PSQt). 
To avoid fine tuning, the lightest \PSQt and the superpartners of the 
Higgs bosons (higgsinos) must have masses near the weak
scale~\cite{Barbieri:1987fn,Dimopoulos:1995mi,Pomarol:1995xc,Cohen:1996vb,Papucci:2011wy,Brust:2011tb}, 
and could therefore have nonnegligible production cross sections at the CERN Large 
Hadron Collider (LHC).

Most searches for the \PSQt look for an excess of events with 
large missing transverse momentum \ptmiss originating from the undetected 
lightest SUSY particle (LSP) produced in \PSQt decays. 
It is typical in these searches to assume that the LSP is the 
lightest neutralino \lsp, which is stable if $R$-parity~\cite{rpv} is conserved. 
However, it has been shown~\cite{Alves:2011sq,Lisanti:2011tm, stealthSusy1} that 
this search strategy is not sensitive to well-motivated 
SUSY models that predict signatures with low \ptmiss in models with gauge mediated SUSY breaking~\cite{giudice}, 
compressed mass spectra~\cite{martin1, martin2}, hidden
valleys~\cite{Strassler:2008bv}, or other mechanisms.  
As searches performed at the LHC using events with high \ptmiss 
set ever more stringent lower bounds on the \PSQt mass~\cite{Sirunyan:2017pjw,Sirunyan:2017wif,Sirunyan:2017xse,Sirunyan:2017dilep,Aaboud:2017ayj,Aaboud:2017aeu}, 
searches for low-\ptmiss alternatives become increasingly 
important.

\begin{figure}[tp]
\centering
\includegraphics[width=0.4\textwidth]{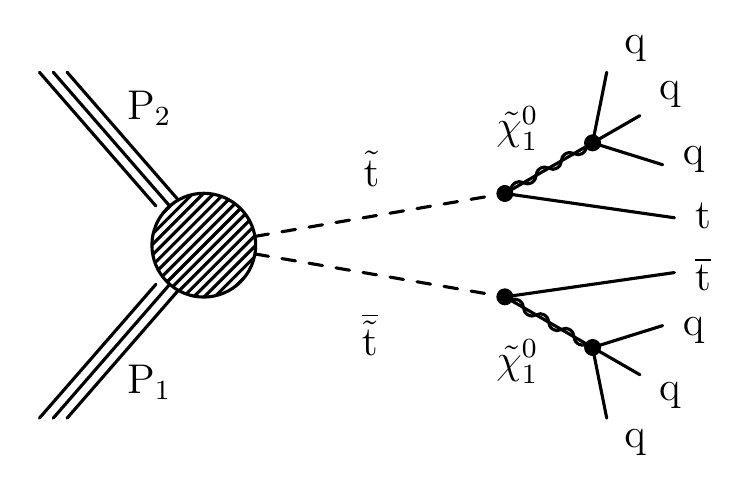}
\includegraphics[width=0.4\textwidth]{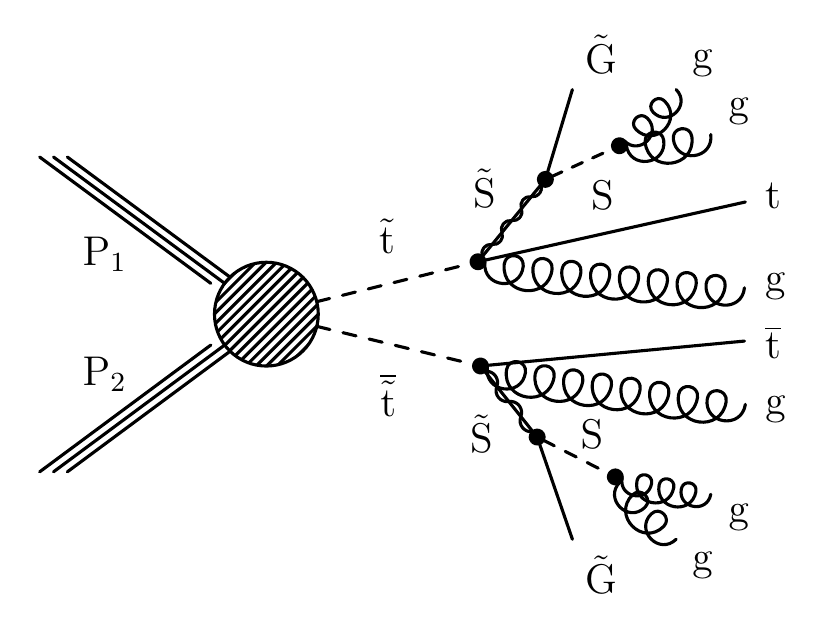}
\caption{ Diagrams of top squark pair production with decays to top quarks and additional 
light-flavor quarks for the RPV SUSY model (\cmsLeft) and with decays to top quarks 
and gluons for the stealth \syy model (\cmsRight).
\label{fig:feynmanDiagrams}}
\end{figure}

Models of $R$-parity violating (RPV) SUSY produce low-\ptmiss signatures 
by providing a mechanism for the LSP, in this case \lsp, to decay.
Among other couplings, RPV SUSY includes a trilinear Yukawa coupling between 
quarks and squarks that allows the \lsp 
to decay into three quarks via an off-shell squark~\cite{rpv}. 
These couplings are typically referred to as $\Lampp_{ijk}$ where $i, j,$ and $k$ 
specify the generations of the participating (s)quarks.
The benchmark RPV model used in this analysis is illustrated 
in Fig.~\ref{fig:feynmanDiagrams}. The \PSQt decays in 
the typical way into a top quark and a \lsp, 
and the \lsp undergoes an RPV decay via 
nonzero $\Lampp_{112}$ into three light-flavor 
quarks, $\lsp\to\PQu\PQd\PQs$. 
However, since this analysis does not distinguish between jets originating 
from quarks of the first and second generation, our results are 
more broadly applicable to any RPV model with coupling $\Lampp_{abc}$ with 
$a,b,c \in \{1,2\}$.

Stealth SUSY models~\cite{stealthSusy1, Fan:2012jf, Fan:2015mxp}  
introduce a new hidden ``stealth'' sector of light particles with 
small or absent couplings to the SUSY breaking sector and 
finite couplings to the visible sector. Because of the weak 
connection to the SUSY breaking sector, SUSY is approximately 
conserved in the stealth sector, resulting in stealth particles 
that are nearly mass-degenerate with their superpartners.  
Production and decay of stealth particles via interactions 
with visible particles can be achieved through a variety 
of ``portals'' including mediation by the Higgs boson or 
new particles at a higher mass scale. 
The benchmark stealth SUSY model used in the interpretation of 
the results of this search (stealth \syy)~\cite{Fan:2015mxp} assumes a minimal 
stealth sector containing only one scalar particle \singlet with 
even $R$-parity and its superpartner \sgo, both of which are 
singlets under all SM interactions, and a portal mediated by 
loop interactions involving a new vector-like messenger field 
(Y), the gluon (\Pg), \lsp, \singlet, and \sgo.
Decays of the \PSQt in the stealth \syy model are illustrated 
in Fig.~\ref{fig:feynmanDiagrams}.  Each \PSQt decays to a 
gluon, top quark, and \sgo, with subsequent decays 
of \sgo to \singlet and a gravitino \PXXSG 
and \singlet to jets via $\singlet \to \Pg\Pg$.
Because of the small mass splitting between the 
\singlet and \sgo, as well as the small \PXXSG mass, the 
undetected \PXXSG carries away very little momentum. 
Thus, the stealth \syy model shares the general feature 
of all stealth SUSY models in that it naturally produces 
a low-\ptmiss signature without $R$-parity violation 
or a special tuning of sparticle masses.

The RPV and stealth \syy models are characterized by the masses of the particles 
and branching fractions in the decay chain. In the benchmark RPV model, we take 
the \lsp mass to be 100\GeV.  For the benchmark stealth \syy model, the critical 
small \sgo-\singlet mass splitting is held constant at 10\GeV, 
and we assume a \sgo mass of 100\GeV and a \PXXSG mass of 1\GeV.  
For both models, a range of \PSQt masses (\mstop) 
are considered from 300 to 1400\GeV, and all decays described 
above are assumed to be prompt with unity branching fractions.

In this paper, we describe a search for \PSQt 
pair production followed by the decay of each \PSQt into 
a top quark and three light-flavor jets via the 
benchmark RPV and stealth \syy models described above.  
This is the first search of its kind at the LHC. 
Previous searches for RPV \PSQt decays focused on final states with 
dijet resonances~\cite{Aaboud:2017nmi, Sirunyan:2018rlj}, 
lepton-jet resonances~\cite{Aaboud:2017opj, Khachatryan:2014ura}, 
intermediate leptonic chargino decays~\cite{Khachatryan:2016ycy},
or final states with many \PQb quarks~\cite{Aaboud:2017faq}.
Previous searches for stealth SUSY targeted superpartners of 
light-flavor quarks with decays into gauge bosons and 
jets~\cite{CMS:2014exa, CMS:2012un}.
Measurements of the \ttbar differential production cross section have been reinterpreted in the context of RPV and stealth SUSY~\cite{Evans:2012bf,Fan:2015mxp} and were found to yield weak constraints for the models considered in this paper.

Before describing each step in more detail in subsequent sections,
we provide an overview of the analysis strategy here. The main distinguishing feature of the signals in 
this analysis, in addition to the presence of two top quarks, is high jet 
multiplicity (\njets).  The SM backgrounds arise through 
processes including top quark pair production (\ttbar), 
multijet production from quantum chromodynamics (QCD), 
production of \ttbar in association with SM weak gauge bosons or additional top quarks ({\ttbar}+X),
production of weak gauge bosons, and single top quark production (other).
These SM processes all include additional jets from initial- and final-state radiation (ISR and FSR).  
The QCD background is primarily suppressed by requiring the presence of 
exactly one charged lepton (\Pe or \Pgm) arising from the leptonic decay of a top quark. 
Backgrounds that do not produce any top quarks are suppressed by 
requiring the presence of at least one jet identified as arising from the 
fragmentation of a bottom quark (\PQb-tagged jet), and additionally that the 
invariant mass of the lepton and a \PQb-tagged jet be consistent with 
the presence of a top quark. 

The signal is distinguished from the dominant and irreducible \ttbar background by means of a 
neural network (NN) trained to recognize differences in the spatial distribution of jets 
and decay kinematic distributions between signal and \ttbar background events.  
Events are divided into 24 categories based on 
their NN score (\snn) and \njets; categories with higher (lower) \snn and \njets tend to be 
signal enriched (depleted).  We perform a simultaneous fit to the number of events in data in 
\snn and \njets categories to estimate the total numbers of \ttbar and potential signal 
events present in the data, as well as the distribution of \ttbar events in \snn 
and \njets categories.  The NN output is designed to have no dependence on \njets, so that 
the \njets distribution of \ttbar events can be constrained in the fit to be the same for 
all \snn categories.  This requirement for \ttbar \njets shape invariance is important 
for the analysis and will be discussed throughout the paper.

This paper is organized as follows.  We introduce the CMS detector and methods for event 
reconstruction and selection in Section~\ref{sec:sele}. Samples of simulated events 
are described in Section~\ref{sec:sim}. The estimation and modeling of SM backgrounds 
are explained in Section~\ref{sec:bkg}, and the description of the treatment of systematic
uncertainties is in Section~\ref{sec:syst}. Finally, the results and their
interpretation are in Section~\ref{sec:res}, followed by the summary in Section~\ref{sec:sum}.

\section{Experimental techniques}
\label{sec:sele}

The search is performed using a data sample of proton-proton ($\Pp\Pp$) collisions 
at $\sqrt{s} = 13\TeV$, corresponding to an integrated luminosity of 137\fbinv,
collected in 2016--2018 with the CMS detector at the LHC.  
Data and simulation samples from four periods (2016, 2017, 2018A, 2018B) are treated 
separately in order to address variations in detector and LHC conditions. Data from 
2018 are divided into two samples (2018A and 2018B), with 2018B corresponding to 
the period when a detector malfunction prevented readout from 3\% of the hadron calorimeter. 
In this section, we define reconstructed physics objects and describe 
the selection criteria for events in the signal 
region (SR) and the control region (CR) of the analysis.

The central feature of the CMS apparatus is a superconducting solenoid 
of 6\unit{m} internal diameter, providing a magnetic field of 3.8\unit{T}. 
Within the solenoid volume are a silicon pixel and strip tracker, a lead 
tungstate crystal electromagnetic calorimeter, and a brass and scintillator 
hadron calorimeter, each composed of a barrel and two endcap sections. 
Forward calorimeters extend the pseudorapidity coverage provided by the barrel 
and endcap detectors. Muons are detected in gas-ionization chambers embedded 
in the steel flux-return yoke outside the solenoid.  A more detailed description 
of the CMS detector, together with a definition of the coordinate 
system used and the relevant kinematic variables, can be found 
in Ref.~\cite{Chatrchyan:2008zzk}.

The CMS trigger system is described in Ref.~\cite{Khachatryan:2016bia}.
Events are selected using triggers that require the presence of at least 
one electron or one muon. The minimum transverse momentum \pt threshold 
is 27 (35)\GeV for electrons and 24 (24)\GeV for muons in 2016 (2017--2018).
The triggers at these thresholds require the lepton to be isolated from 
tracks and calorimeter deposits in the detector.  
Events may also be selected from single-lepton triggers with higher \pt thresholds,
115\GeV for electrons and 50\GeV for muons, with no isolation requirements.
The combined trigger efficiency varies from 80\% for leptons with \pt 
close to the lower thresholds to greater than 95\% for leptons with 
$\pt > 120\GeV$.

Events are reconstructed using the particle-flow (PF) algorithm~\cite{CMS-PRF-14-001}, 
which reconstructs particles in an event using an optimized combination of 
information from the various elements of the CMS detector and identifies each 
as a photon, electron, muon, charged hadron, or neutral hadron.  These particles 
are further clustered into jets as described below.  

The reconstructed vertex with the largest value of summed physics-object $\ensuremath{p_{\mathrm{T}}}^2$ 
is taken to be the primary $\Pp\Pp$ interaction vertex, where the physics objects are the 
jets, clustered using the anti-\kt algorithm~\cite{Cacciari:2008gp, Cacciari:2011ma} 
with the charged-particle tracks assigned to the vertex as inputs, and the 
associated missing transverse momentum, taken as the negative vector sum of the \pt of those jets~\cite{CMS-TDR-15-02}.
Charged-particle tracks associated with vertices from other $\Pp\Pp$ interactions (pileup) 
are removed from further consideration.  The primary vertex is required to lie within 
24\unit{cm} of the interaction point along the beam axis, and within 2\unit{cm} in the plane 
transverse to the beam axis.

Electrons and muons must satisfy $\pt > 30\GeV$ and $\abs{\eta} < 2.4$. 
For the analysis of the 2017 and 2018 data, the electron \pt threshold is 
increased to 37\GeV to account for the higher trigger threshold. 
The lepton identification requirements are the 
``tight'' criteria for electrons~\cite{electronID} and the ``medium'' 
criteria for muons~\cite{muonID}.
Leptons must be isolated within a cone of radius 
$R = \sqrt{\smash[b]{ (\Delta \phi)^2 + (\Delta \eta)^2}}$
 that scales as $1/\pt$ between a maximum of 0.2 for leptons with
 $\pt < 50\GeV$ and a minimum of 0.05 for lepton $\pt > 200\GeV$~\cite{miniIso}.

Jets are clustered from the reconstructed PF particles using the 
anti-\kt algorithm with a distance parameter of 0.4.  Criteria are 
applied to remove events with jets arising from instrumental effects 
or reconstruction failures~\cite{jetID,CMS-PAS-JME-16-003}.  The reconstructed jet energies 
are corrected for the nonlinear response of the detector~\cite{Khachatryan:2016kdb,CMS-DP-2020-019} and 
for contributions from neutral hadrons from pileup~\cite{Cacciari:2007fd}.
Jets are required to have $\pt>30\GeV$ and $\abs{\eta}<2.4$.  
Jets overlapping with a selected lepton within a cone of radius 
$R=0.4$ are removed.  A neural network-based algorithm~\cite{Sirunyan:2017ezt} 
is used to identify \PQb quark jets; for jets with $\pt$ around $30\GeV$, the algorithm 
has an efficiency of 65\% and a misidentification rate for light-flavor 
jets (including gluon jets) of 1\%.

In addition to the trigger and vertex criteria above, events in the 
SR must contain exactly one isolated electron or muon and at least seven jets, 
at least one of which should be \PQb tagged. Samples with seven and 
eight jets include a small number of expected signal events, 
but are included in the SR to constrain the background. To further 
reject the QCD background, we require the scalar sum of jet \pt (\HT) 
to exceed 300\GeV. To suppress non-\ttbar backgrounds, we require the 
invariant mass of the system formed by the \PQb-tagged jet and the lepton 
to be between 50 and 250\GeV. If there is more than one \PQb-tagged jet 
in the event, the invariant mass of each \PQb-tagged jet and the lepton 
is considered, and at least one combination is required to meet the 
above criterion. No requirement is made on the event \ptmiss.

In addition to the SR, a signal-depleted control region (CR) 
dominated by QCD background is defined with the dual 
purpose of determining the QCD contribution to the SR 
and verifying the important assumption of \ttbar \njets 
shape invariance with \snn.  Despite being dominated by 
QCD background, the CR is useful for confirming 
\ttbar \njets shape invariance because many of the 
jets used as inputs to the NN arise from QCD radiation, 
which is common to the \ttbar and QCD backgrounds; 
this claim is verified in Section~\ref{sec:syst}.
The CR is defined similarly to the SR with the 
differences being that the lepton is required to be a muon;
the muon is required to fail the SR isolation requirement;
there is no requirement for a \PQb-tagged jet, 
nor on the invariant mass of the lepton and \PQb-tagged jet;
the only trigger used is the high-threshold muon trigger without an isolation 
requirement; and the muon \pt threshold is 55\GeV.

\section{Simulated event samples}
\label{sec:sim}

Simulated event samples are used in the estimation of the 
expected number of SM background and signal events passing the SR selection.
Top quark pair and single top quark events produced in the $t$~channel 
are generated with the next-to-leading-order (NLO) 
\POWHEG v2.0~\cite{Nason:2004rx,Frixione:2007vw,Alioli:2010xd,Frixione:2007nw,Frederix:2012dh} 
generator, while single top quark events in the t\PW~channel are generated 
with \POWHEG v1.0~\cite{Frixione:2007nw}. 
Single top quark production in the $s$~channel, 
as well as rare SM processes such as $\ttbar\cPZ$ and $\ttbar\PW$ 
are generated at NLO accuracy with the \MGvATNLO v2.2.2 program. 
The \MGvATNLO v2.2.2 generator~\cite{Alwall:2014hca,Kalogeropoulos:2018cke} is 
used in the leading-order (LO) mode to simulate QCD and ${\PW}+$jets events.

For the signal, top squark pair production events are generated 
using \MGvATNLO in LO mode, including up to two additional partons 
in the matrix element calculation. The top squarks are decayed 
using \PYTHIA v8.212 (2016) or 8.226 (2017--2018)~\cite{pythia8}
according to the signal models described in Section~\ref{sec:intro}.  
The signal production cross section ($\sigma_{\PSQt\PASQt}$) is calculated 
as a function of \mstop using approximate next-to-NLO (NNLO) plus 
next-to-next-to-leading-logarithm (NNLL) 
calculations~\cite{Borschensky:2014cia,Beenakker:2016lwe}. 

The generation of these processes is based on either LO or NLO parton distribution 
functions (PDFs) using NNPDF3.0~\cite{Ball:2014uwa} for the simulated samples 
corresponding to 2016 detector conditions, and using the NNLO PDF sets from 
NNPDF3.1~\cite{Ball:2017nwa} for the 2017 and 2018 simulated samples. 
Parton showering and hadronization are simulated with {\PYTHIA}
using underlying event tune CUETP8M1~\cite{Khachatryan:2015pea} for 2016 samples, 
except for \ttbar production which used tune CUETP8M2T4~\cite{CMS-PAS-TOP-16-021}, 
or {\PYTHIA} with tune CP5 (CP2)~\cite{Sirunyan:2019dfx} for all 
2017 and 2018 background (signal) samples. To model the effects of pileup, 
simulated events are generated with a nominal distribution of pp interactions 
per bunch crossing and then reweighted to match the corresponding distribution in data.
The CMS detector response is simulated using a \GEANTfour-based 
model~\cite{Agostinelli:2002hh}, and event reconstruction is performed in the 
same manner as for collision data.  The most precise cross section 
calculations available are used to normalize the SM simulated samples, 
corresponding to NLO or NNLO accuracy in most cases~\cite{Alwall:2014hca,
Czakon:2011xx,Kant:2014oha,Aliev:2010zk,Gehrmann:2014fva,Campbell:1999ah,
Campbell:2011bn,Li:2012wna}.  

The simulation is corrected to eliminate small discrepancies between 
data and simulation in the trigger efficiency, lepton selection efficiency, and
$\PQb$ tagging efficiency.  Analysis-specific corrections for the \HT distribution in 
\ttbar simulation, parameterized as functions of \njets and \HT, are 
obtained in a signal-depleted sample identical to the SR, 
except for the requirement $5 \leq \njets \leq 7$.  
Events with $\njets = 7$ are common to the SR, but as mentioned above,
this sample has low signal contamination.
The correction is parameterized with an exponential function in \HT 
with parameters depending linearly on \njets in order to extend the 
correction into the $\njets > 7$ SR.  The \HT correction is small 
at low \HT and 20--40\% at $\HT=1500\GeV$, depending on \njets.

\section{Background estimation}
\label{sec:bkg} 

Simulated background events passing the SR selection requirements 
predominantly originate from \ttbar production, with 
contributions of less than 10\% from QCD, and a few 
percent from the remaining minor backgrounds including 
\ttbar production in association with a vector boson, 
single top quark production, and ${\PW}+$jets.

As introduced in Section~\ref{sec:intro}, the crux of the 
analysis is to estimate the dominant \ttbar background 
in four bins of \snn and six \njets bins using a simultaneous 
binned maximum-likelihood fit constraining the \ttbar 
\njets shape to be the same in all \snn categories.  
Event yields, as well as the \njets and \snn distributions, are fixed at 
values determined from a signal-depleted data control sample for 
the QCD background and from simulation for the minor backgrounds, as 
described later in this section. The yield and \njets shape of the \ttbar
background, along with the signal strength, are 
determined in the fit; signal strength is defined as 
the ratio of the fit signal event yield to the one predicted by SUSY.
  
The NN is trained to discriminate between signal and \ttbar background by 
exploiting differences in the event shape and distributions of the kinematic variables.  
The gradient reversal technique~\cite{Ganin2015} is used to minimize
dependence of the NN output on \njets, as required by the 
primary assumption that the \ttbar \njets shape is the same in 
all \snn categories.  All NN input variables are computed in an 
approximate center-of-mass frame defined by all jets in 
the event with $\pt>30\GeV$ and $\abs{\eta}<5$.
The NN input variables include the four-vector components for the 
seven jets in the event with the highest momentum in the center-of-mass frame, 
the four-vector components of the lepton in the event,
the second through fifth Fox--Wolfram moments~\cite{FWmoments} 
normalized by the first moment, and the three eigenvalues of 
the sphericity tensor~\cite{sphericity} normalized by the 
sum of the eigenvalues. The Fox--Wolfram moments and sphericity 
tensor eigenvalues, which are computed from the same 
seven highest momentum jets, quantify the distribution of jet energy in 
the event, which tends to be more spherical for signal \PSQt pair production
than for the \ttbar background. 

\begin{figure}[tbhp]
\centering
\includegraphics[width=0.49\textwidth]{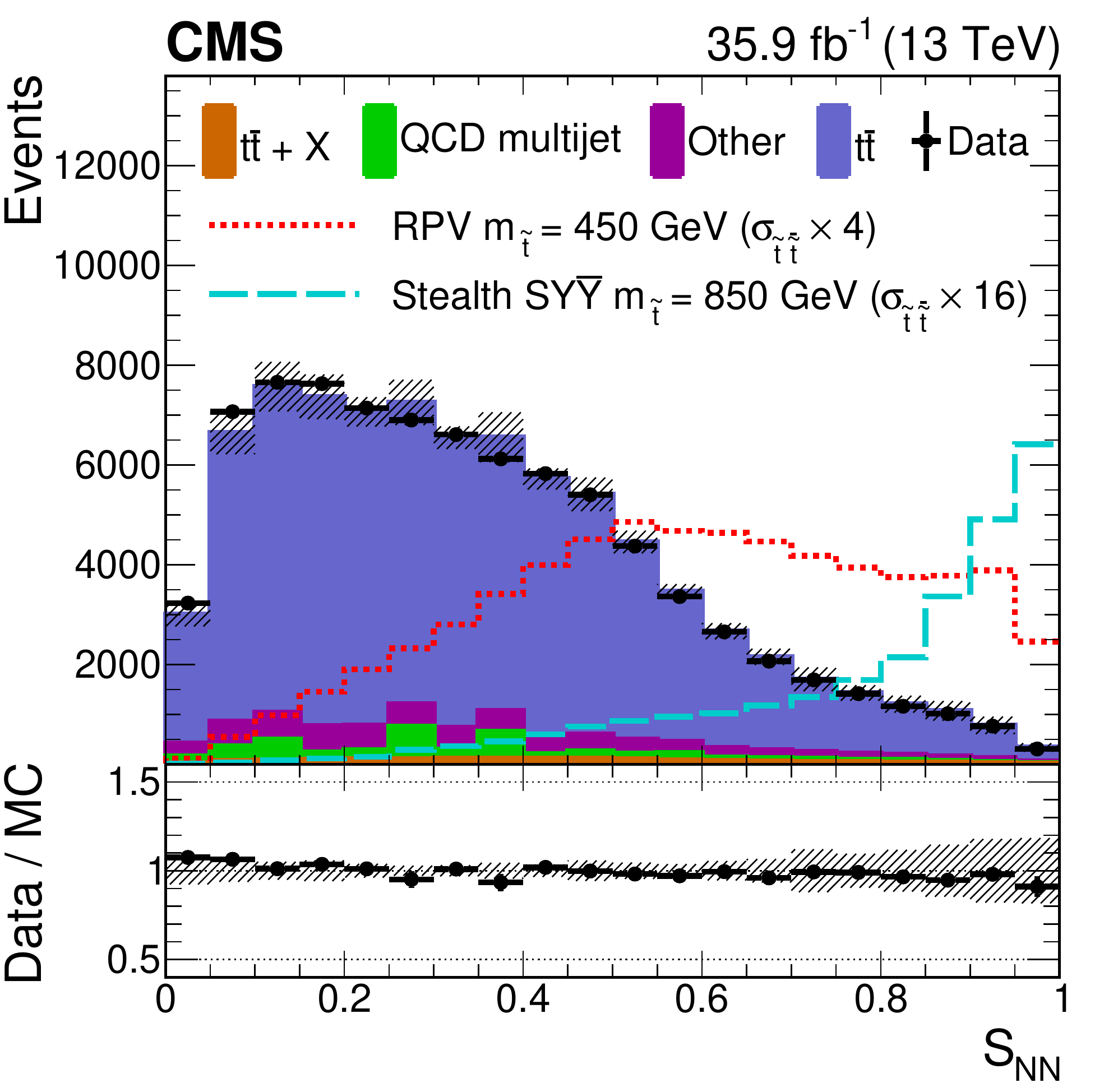}
\includegraphics[width=0.49\textwidth]{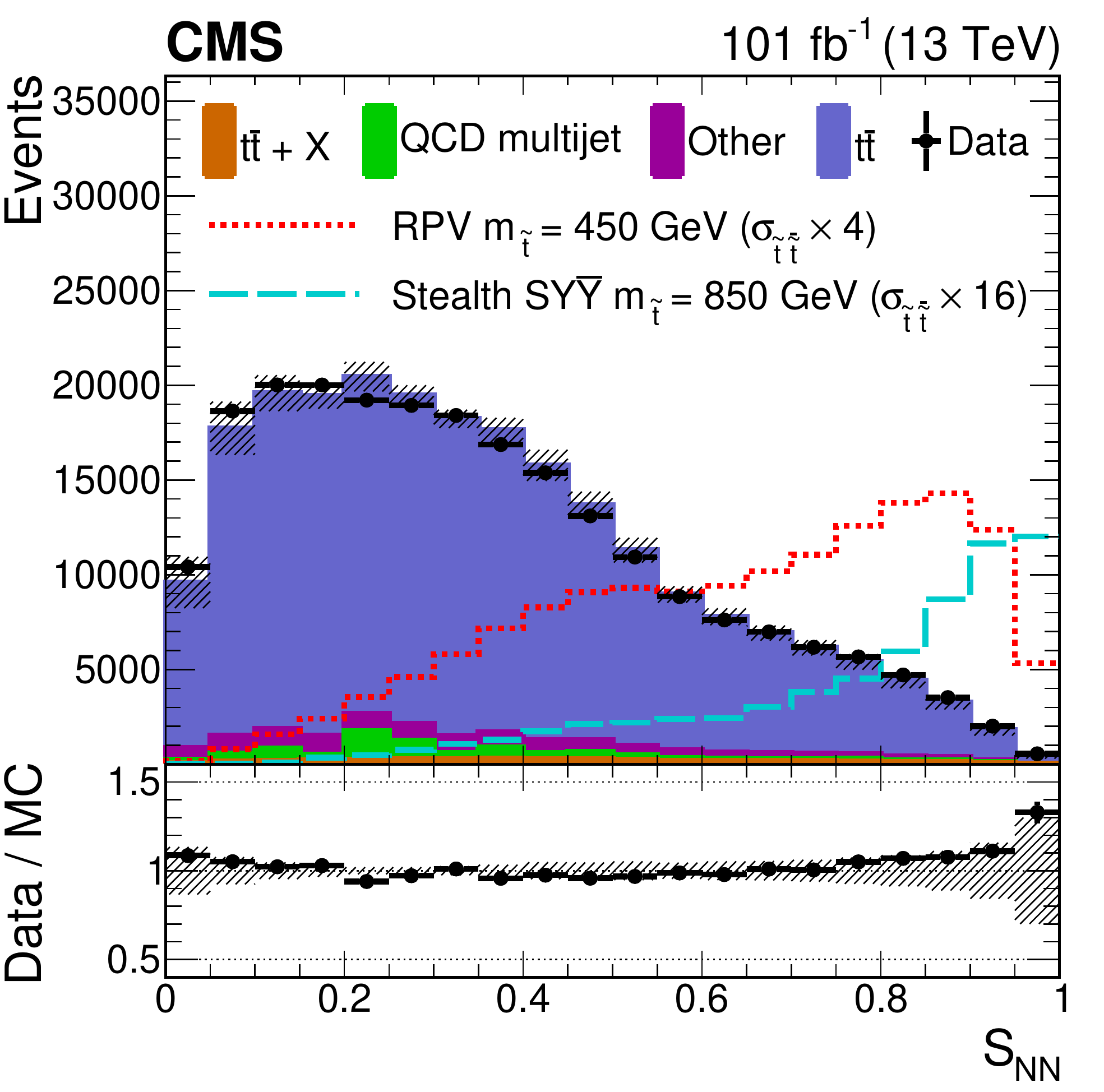}
\caption{ The \snn distributions for the 2016 training (\cmsLeft) and 2017+2018 training (\cmsRight) 
show the corresponding data in the SR (black points);
simulated background normalized to the number of data events (filled histograms);
RPV signal model with \mstop of 450\GeV (red short dashed); 
and stealth \syy signal model with \mstop of 850\GeV (cyan long dashed).
All events shown pass the SR event selection. The band on the total background 
histogram denotes the dominant systematic uncertainties related to the 
modeling of \HT, jet mass, and jet \pt in the \ttbar simulation, 
as well as the statistical uncertainty for the non-\ttbar components.
The lower panel shows the ratio of the number of data events to the 
number of normalized simulated events with the band 
representing the difference between the nominal ratio 
and the ratio obtained when varying the total background 
by its uncertainty. 
\label{fig:nn-distributions}}
\end{figure}

For the NN training, simulated \ttbar events are used for the background 
sample, and a mixture of RPV and stealth \syy simulated events 
with \mstop from 350--850\GeV is used as the signal sample.
In this way, the NN can identify common features among all signal 
samples ensuring a search with broad sensitivity. Reflecting 
differences in simulation between the data taking periods, as described in Section~\ref{sec:sim}, 
a single training is used for 2017, 2018A, and 2018B,
with a separate training used for 2016.
The \snn distributions for the simulated background, several signal models, and 
the 2016 and 2017+2018 data are shown in Fig.~\ref{fig:nn-distributions}. 

For each of the six \njets bins, events are divided into four \snn bins: 
\snni{1} (lowest \snn), \ldots, \snni{4} (highest \snn).  
The \snn bin boundaries are chosen separately for each \njets bin
such that the expected significance for the 550\GeV RPV signal model, 
which has expected significance close to 5 standard deviations ($\sigma$), 
is maximized, under the constraint that the fraction of 
simulated \ttbar events in each \snn bin is the same for all \njets bins.
For example, the fraction of all events in each \njets bin 
falling into the \snni{1} bin is constrained to be approximately $56\%$, 
while the fraction of events falling into the \snni{4} bin is constrained 
to be approximately 2.4\%.
This constraint removes the small dependence of \snn on \njets 
that remains after NN training with gradient reversal.

In the maximum-likelihood fit, the \ttbar \njets distribution is parameterized with 
a function inspired by QCD jet scaling patterns~\cite{Gerwick:2012hq} in which 
the ratio $R(i) = M_{i+1}/M_i$, where $M_i$ is the number of events with $\njets=i$, 
can be described by a falling ``Poisson'' component at low \njets and a 
constant ``staircase'' component at high \njets.
This ratio is well modeled by the function 
\begin{linenomath*}
\begin{equation*}
f(i)  =  a_2 + \left[ \frac{ (a_1 - a_2 )^{i-7} }{ (a_0 - a_2)^{i-9} } \right]^{1/2}.
\end{equation*}
\end{linenomath*}
Notice that $a_0 = f(7)$, $a_1 = f(9)$, and $a_2$ is the asymptotic value for large $i$.  
This particular parameterization was chosen to avoid large correlations between the
fit parameters.  The \njets distribution for each \snn bin $j$ (see Fig.~\ref{fig:results_fitPlots}) 
is modeled using a recursive expression given by $M^j_i = Y^j_7 \, \Pi_{k=7}^{i-1} f(k)$ 
where $Y^j_7$ are normalization parameters that are floating in the fit.  
The last \njets bin considered is an inclusive $\njets \geq 12$ bin, such 
that $i \in [7,12]$. In the maximum-likelihood fit, the free parameters consist 
of the three shape parameters $a_0$, $a_1$, and $a_2$; 
the four normalization parameters $Y^j_7$; the signal strength; 
and all nuisance parameters related to systematic uncertainties 
described in Section~\ref{sec:syst}.

The QCD background yield parameters are fixed in the fit at the values
determined from the CR. More specifically, the QCD background 
prediction for each \njets-\snn bin in the SR is given by the 
yield for the same bin in the CR in data, after subtraction of
the non-QCD backgrounds as predicted from simulation,
multiplied by the ratio of SR to CR yields in simulation ($R_\text{QCD}$).
This procedure is verified with a closure test in the simulation.
The yield parameters from the minor backgrounds are 
also kept fixed in the fit at the values predicted by simulation.  
While the yield parameters are fixed in the fit, the ultimate 
contributions from QCD and minor backgrounds vary according 
to the constrained nuisance parameters related to systematic 
uncertainties in those fit components.

\section{Systematic uncertainties and fit validation}
\label{sec:syst}

As described in Section~\ref{sec:bkg}, an unbiased estimate of 
the dominant \ttbar background is obtained from the fit to 
data as long as the \ttbar \njets shape is the same for 
all four \snn bins.  By construction, 
\njets shape invariance is achieved in the simulation with 
an \njets-specific \snn binning as described 
in the previous section.
Thus, systematic uncertainties on the \ttbar background are 
important to the degree that they violate the assumption that 
the \snn binning determined in simulation also applies to the data.
We quantify how each source of uncertainty causes deviations 
from the assumed \njets shape invariance by comparing 
the nominal \njets shape to the \njets shapes in all \snn 
bins after performing the relevant systematic variation 
to the \ttbar simulation.  Each systematic variation is associated with a constrained nuisance parameter in the fit.
The deviation in shape for each \njets distribution, derived from the ratio of the post-variation shape divided by the nominal shape, changes linearly with the associated nuisance parameter for the systematic variation, while preserving the normalization of the distribution.  

Sources of \ttbar shape uncertainty include uncertainty
in aspects of event generation including PDFs,
choice of renormalization and factorization scales 
($\mu_\mathrm{R}$, $\mu_\mathrm{F}$ scales), and parton shower modeling, 
which is itself composed of aspects related to modeling of
ISR, FSR, color reconnection in the parton shower, 
matrix element-parton shower matching scale (ME-PS), 
underlying event (UE tune), and pileup modeling.
The uncertainty due to the choice in ($\mu_\mathrm{R}$, $\mu_\mathrm{F}$) scales
is determined by independently varying both by factors of 2.0 and 0.5 
excluding the variations (2.0, 0.5) and (0.5, 2.0)~\cite{Kalogeropoulos:2018cke, Cacciari:2003fi, Catani:2003zt}. 
The ISR and FSR uncertainties originate from variations of the renormalization scale for the parton shower by factors 0.5 and 2.0, effectively varying the value of $\alpha_{\text{S}}$.
The color reconnection uncertainty is calculated by allowing resonant decays to occur before the merging of multi-parton systems.
The ME-PS uncertainty is obtained by varying the \POWHEG parameter that 
governs ME-PS matching about its nominal value according to 
$h_\text{damp} = 1.379^{+0.926}_{-0.505}$ times the top quark mass~\cite{Sirunyan:2019dfx}.
The UE tune uncertainty comes from variation of the \PYTHIA 
parameters that control the modeling of the underlying event 
as described in Ref.~\cite{Sirunyan:2019dfx}.
The total inelastic $\Pp\Pp$ cross section is changed by 5\% to 
estimate the uncertainty related to pileup~\cite{Sirunyan:2018nqx}.

Sources of \ttbar shape uncertainty related mostly to 
aspects of detector simulation include determination of 
jet energy scale (JES) and resolution (JER), 
modeling of the \PQb tagging efficiency, 
modeling of the efficiency for lepton triggers, 
identification, and isolation (lepton efficiencies); 
residual mismodeling of \HT, jet \pt, and jet mass; 
and use of the CR for measuring deviations 
from the assumption of \njets shape invariance.

The uncertainty in the modeling of \HT in the \ttbar simulation 
is composed of four separate components. 
The first \HT uncertainty (primary) is taken as the full difference 
in the \ttbar background shape with and without the \HT correction.
The second \HT uncertainty (validation) is taken as the difference 
between the simulation with nominal \HT correction (described in Section~\ref{sec:sim})
and the observed \HT distribution in the signal-depleted SR sample with $\njets=8$.
The third and fourth \HT uncertainties address the choices of 
parameterization of the \HT correction as functions of \HT and \njets.  
For these, we take the uncertainty as the difference between 
the nominal correction and two alternate corrections that 
use the $\HT=2000\GeV$ correction for all events 
with $\HT>2000\GeV$ (\HT-parameterization) and
the $\njets=7$ correction for all values of 
\njets (\njets-parameterization).

Comparisons of data and simulation in the CR show that the simulation
predicts distributions with higher values of jet \pt and mass than 
observed. The observed discrepancy at jet \pt (mass) of 
400 (50)\GeV depends on jet \pt rank and is small for the 
highest \pt jet in each event growing to approximately 50\% for 
the jet with sixth-highest \pt in each event. 
Similar trends are observed
in the signal-depleted, \ttbar-dominated SR with $\njets=7$.
In the CR, the discrepancy in the falling tail of each 
distribution is minimized when the \pt (mass) of each jet is 
scaled by the value 0.95, 0.95, 0.95, 0.95 (0.95, 1.01, 0.98, 0.98)
for 2016, 2017, 2018A, and 2018B, respectively.  
Thus, the related \ttbar shape uncertainty is taken to be the resulting 
difference between scaled and nominal simulated \ttbar distributions.
The dependence on jet \pt rank indicates 
that the discrepancy arises predominantly in the event generation; however, we
choose to estimate the associated systematic uncertainty separately for each
data taking period to include potential effects of detector response simulation.  
The \HT correction is omitted from the determination of 
these jet \pt and mass uncertainties to avoid double counting of 
\HT mismodeling effects. 
In addition, because the estimation of jet \pt and mass 
uncertainties relies on variable scaling (rather than event 
reweighting), the uncertainties include effects of changes in the \snn for 
each event, which is not included in the \HT uncertainty.

\begin{figure}[tbp]
\centering
\includegraphics[width=0.47\textwidth]{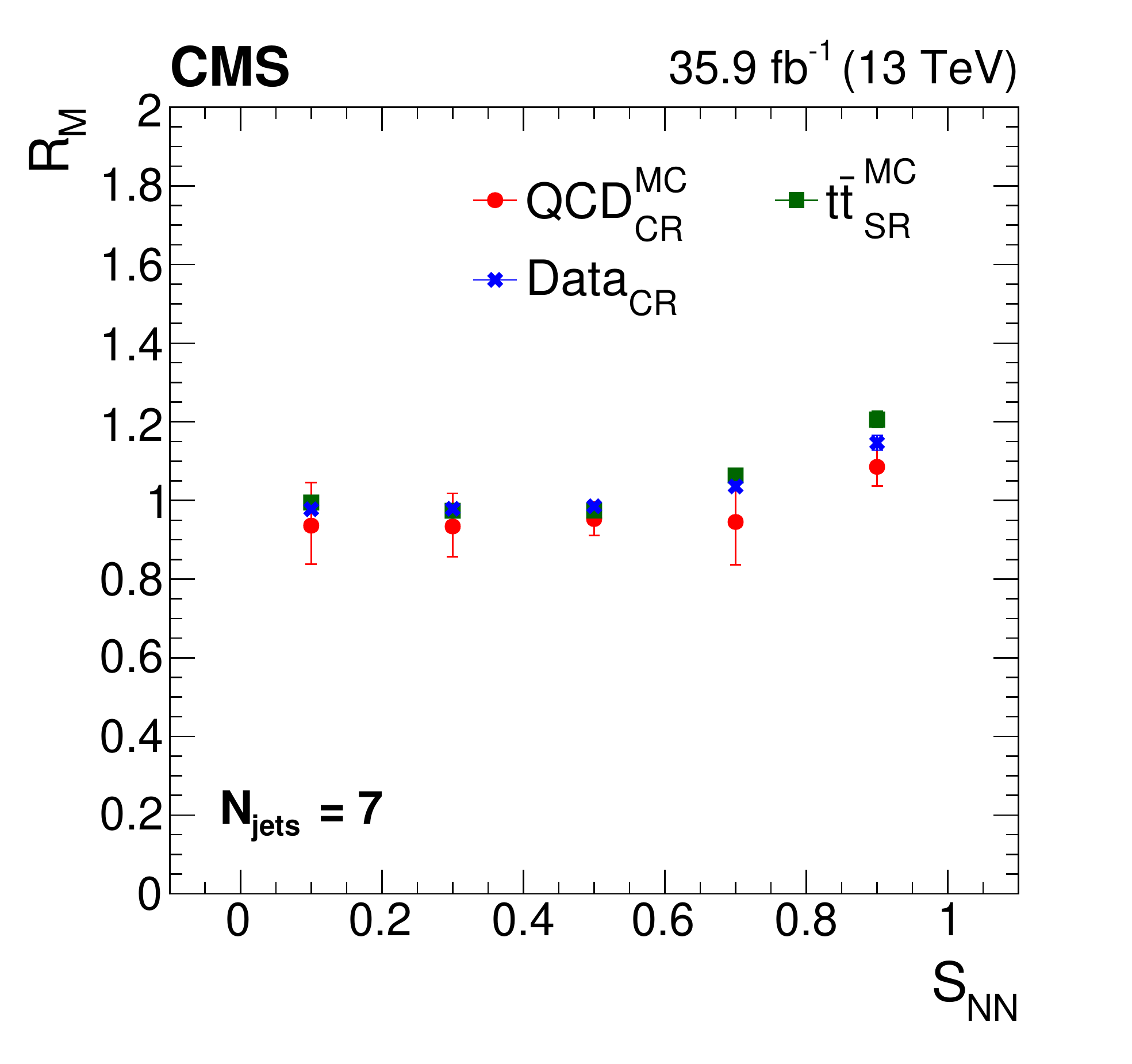}
\includegraphics[width=0.47\textwidth]{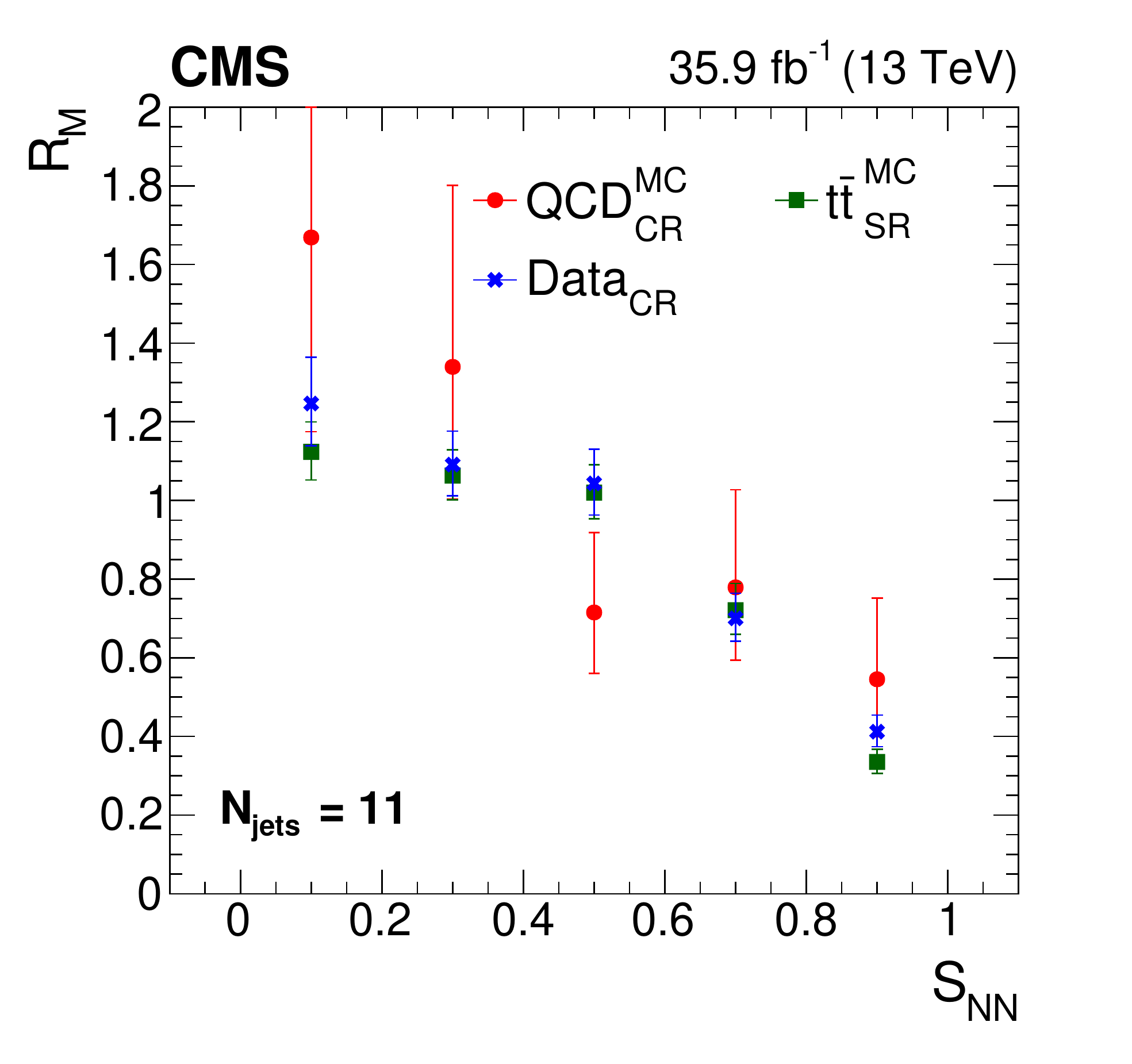}
\caption{Distribution in \snn of the ratio $R_M$, as defined in the text, for $\njets = 7$ (\cmsLeft) and 11 (\cmsRight), for the QCD CR simulation (red circles), the \ttbar SR simulation (green squares), and data in the CR (blue crosses) for the 2016 data period. The error bars indicate the statistical uncertainty in the value of $R_M$. 
\label{fig:qcdcrestimate}}
\end{figure}

As mentioned above, the use of \njets-dependent \snn binning ensures 
that the \njets shape is the same in all four \snn bins in simulation, 
and the use of the same binning in the data assumes that the 
\njets-\snn dependence is well modeled in the simulation. 
This assumption is confirmed and a related systematic uncertainty 
is determined by comparing the \njets shapes (in five uniform \snn bins)
for data and simulation in the CR.
For each of the six \njets bins, we compute the ratio
$R_M = (1/\mu_i)\,(M_\text{all}/M_i)$ as a function of \snn,
where $M_\text{all}$ is the total number of events in all \njets bins,
$M_i$ is the total number of events in the $\njets=i$ bin,
and $\mu_i$ is the uncertainty-weighted average of $M_\text{all}/M_i$ in 
the $\njets=i$ bin used to facilitate comparison of the $R_M$ shapes 
between samples and \njets bins with different normalizations.
Figure~\ref{fig:qcdcrestimate} shows a comparison of $R_M$ (from $\njets=7$ and 11 
in the 2016 analysis) for simulation of the QCD background in the CR ($\mathrm{QCD_{CR}^{MC}}$), 
simulation of \ttbar in the SR ($\mathrm{\ttbar_{ SR}^{ MC}}$), and the data in the 
QCD background-dominated CR ($\mathrm{Data_{CR}}$). 
Agreement between $\mathrm{QCD_{CR}^{MC}}$ and $\mathrm{\ttbar_{ SR}^{ MC}}$ 
demonstrates that QCD background-dominated data in the CR are a good proxy for 
\ttbar-dominated data in the SR, and agreement between $\mathrm{QCD_{CR}^{MC}}$ and 
$\mathrm{Data_{CR}}$ verifies that the dependence of the \njets shape on \snn is well 
modeled in the simulation. Similar agreement is found for the $R_M$ distributions 
for the other \njets bins and data periods. The uncertainty related to the 
combination of both effects is taken as the difference between 
$\mathrm{\ttbar_{ SR}^{ MC}}$ and $\mathrm{Data_{CR}}$.

For the QCD background, the shape is obtained from data in the CR, 
and the normalization is set with $R_\text{QCD}$.  Because the systematic 
uncertainties in the simulation largely cancel in the $R_\text{QCD}$ ratio, 
the uncertainty in $R_\text{QCD}$ is dominated by the statistical uncertainty 
of simulated samples and ranges from 15--25\% depending on data collection period.

Sources of systematic uncertainty in the predictions for 
signals and the minor 
backgrounds include PDFs, JES, JER, \PQb tagging efficiency, 
lepton efficiency, trigger efficiency, ($\mu_\mathrm{R}$, $\mu_\mathrm{F}$) scales, 
cross sections for the minor backgrounds, 
and a 2.3--2.5\% uncertainty in the integrated luminosity~\cite{lumi16, lumi17, lumi18}. 
Since the signal and minor backgrounds are estimated directly 
from simulation, related uncertainties are included as the full 
effect of the systematic variation on the yields in each \njets 
and \snn bin, thereby taking into account normalization effects as well as shape changes.

Uncertainties derived from comparisons of data and 
simulation separately in each data taking period (related to pileup, 
JES, JER, \PQb tagging efficiency, lepton efficiencies, 
\HT corrections, \njets shape invariance, and integrated luminosity) 
are treated as uncorrelated among all data samples.
Uncertainties related to parton shower modeling 
are treated as fully correlated for 2017, 2018A, and 2018B, 
while the corresponding uncertainties for 2016 are uncorrelated with those from the other data taking periods;  
uncertainties related to  ($\mu_\mathrm{R}$, $\mu_\mathrm{F}$) scales and 
cross sections for the minor backgrounds are treated as 
correlated between all four periods.

Table~\ref{tab:systematics} shows the impact of the 
systematic uncertainties on the expected event yields 
for the \ttbar background, minor backgrounds, and 
the RPV signal model with $\mstop=550\GeV$.
For sources of uncertainty for which the size of the impact
depends on \njets and \snn, a representative range of values 
is listed along with the maximum value from all bins.

\begin{table}[htbp]
\centering
\topcaption{Summary of the impact of systematic uncertainties in the 
expected event yields for the \ttbar background, 
minor backgrounds (both {\ttbar}+X and other), and the RPV signal model with $\mstop=550\GeV$.
Abbreviated names for each source of uncertainty are explained in the text.
For sources of uncertainty for which the size of the impact 
depends on \njets and \snn, a representative 
range of values showing the 
16th and 84th percentile of 
all the corrections is listed with the maximum value from all 
bins shown in parentheses.  All values are in units of percent.
}
\begin{scotch}{l r r r}
\multirow{2}{*}       & \ttbar     & Minor      & RPV    \\
Source of uncertainty & background & background & signal \\ 
\hline
PDFs                   & 0--1   (2) & 0--1    (8) &  0--2  (7) \\ 
($\mu_\mathrm{R}$, $\mu_\mathrm{F}$) scales & 0--2   (5) & 1--8   (18) &  0--3  (4) \\ 
ISR                    & 0--4  (15) & \NA & \NA \\ 
FSR                    & 0--8  (27) & \NA & \NA \\ 
Color reconnection     & 0--10 (44) & \NA & \NA \\ 
ME-PS                  & 0--14 (82) & \NA & \NA \\ 
UE tune                & 0--7 (100) & \NA & \NA \\ 
Pileup                 & 0--2   (7) & 0--7   (28) &  0--2  (4) \\[\cmsTabSkip]

JES                       & 0--4 (18) & 5--21 (100) & 1--11 (31) \\ 
JER                       & 0--2 (10) & 1--15 (100) &  0--6 (14) \\ 
\PQb tagging             & 0--1  (3) & 0--2   (12) &  0--2  (2) \\ 
Lepton efficiencies       & 0--1  (1) & 3--5    (5) &  3--4  (4) \\[\cmsTabSkip] 

\HT primary                 & 0--5  (17) & \NA & \NA \\ 
\HT validation              & 0--1   (4) & 0--6 (10) & \NA \\ 
\HT \HT-parameterization    & 0--2   (9) & \NA & \NA \\ 
\HT \njets-parameterization & 0--7  (27) & \NA & \NA \\ 
Jet \pt                     & 0--4  (15) & \NA & \NA \\ 
Jet mass                    & 0--4  (15) & \NA & \NA \\ 
\njets shape invariance     & 0--12 (37) & \NA & \NA \\[\cmsTabSkip] 

Integrated luminosity       & \NA         & 2.3--2.5         & 2.3--2.5       \\
Theoretical cross section   & \NA         & 30               & \NA         \\
\end{scotch}
\label{tab:systematics}
\end{table}

\section{Results and interpretation}
\label{sec:res}

\begin{figure*}[htp]
\centering
\includegraphics[width=0.82\textwidth]{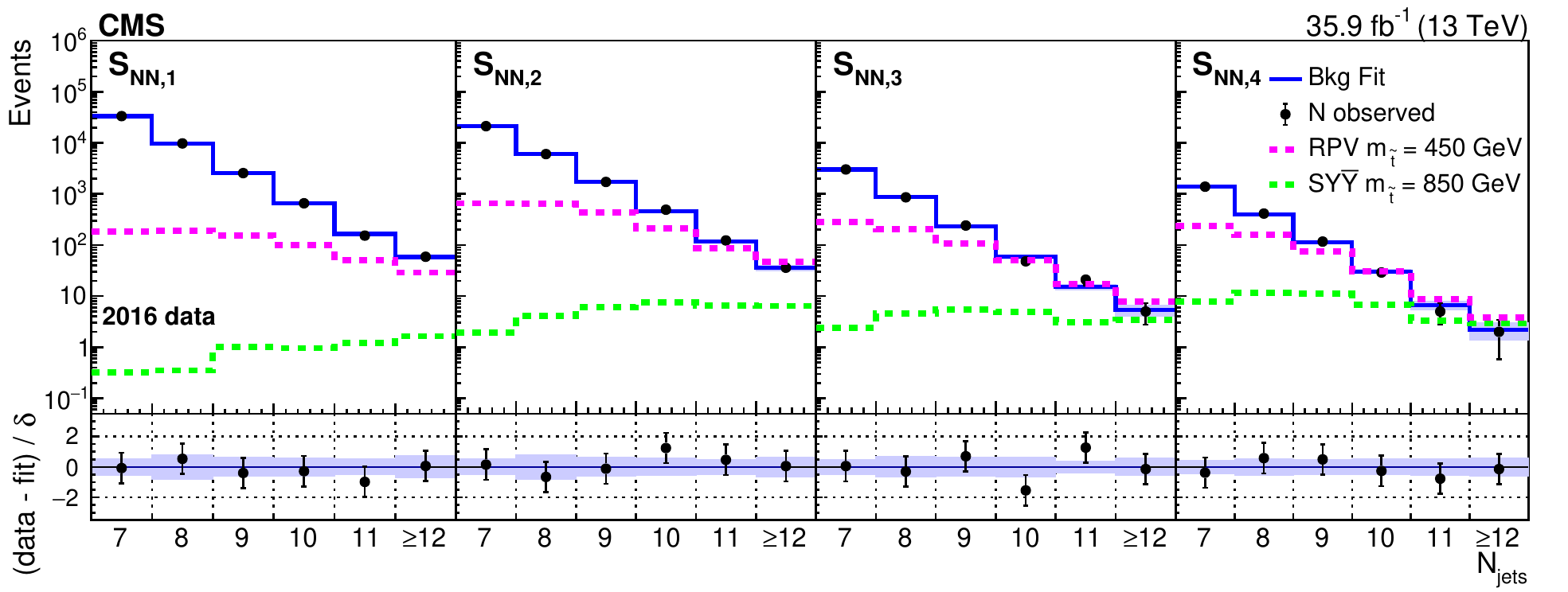}
\includegraphics[width=0.82\textwidth]{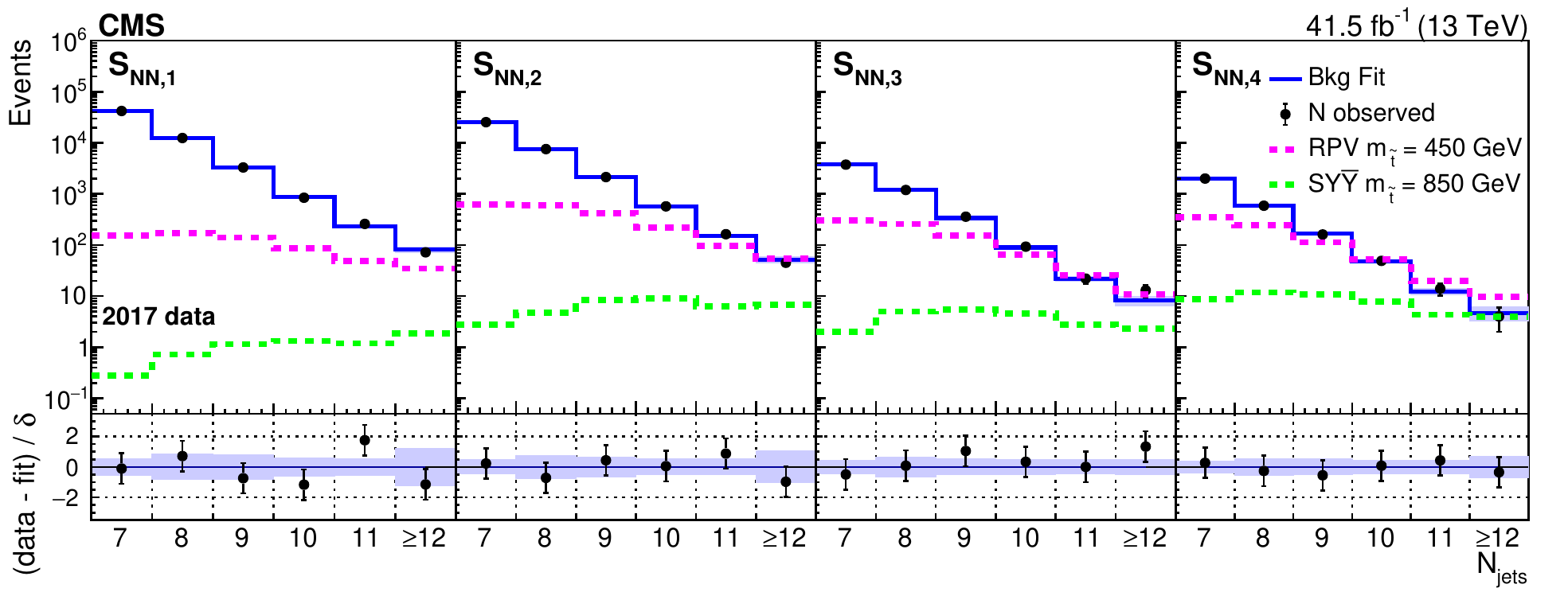}
\includegraphics[width=0.82\textwidth]{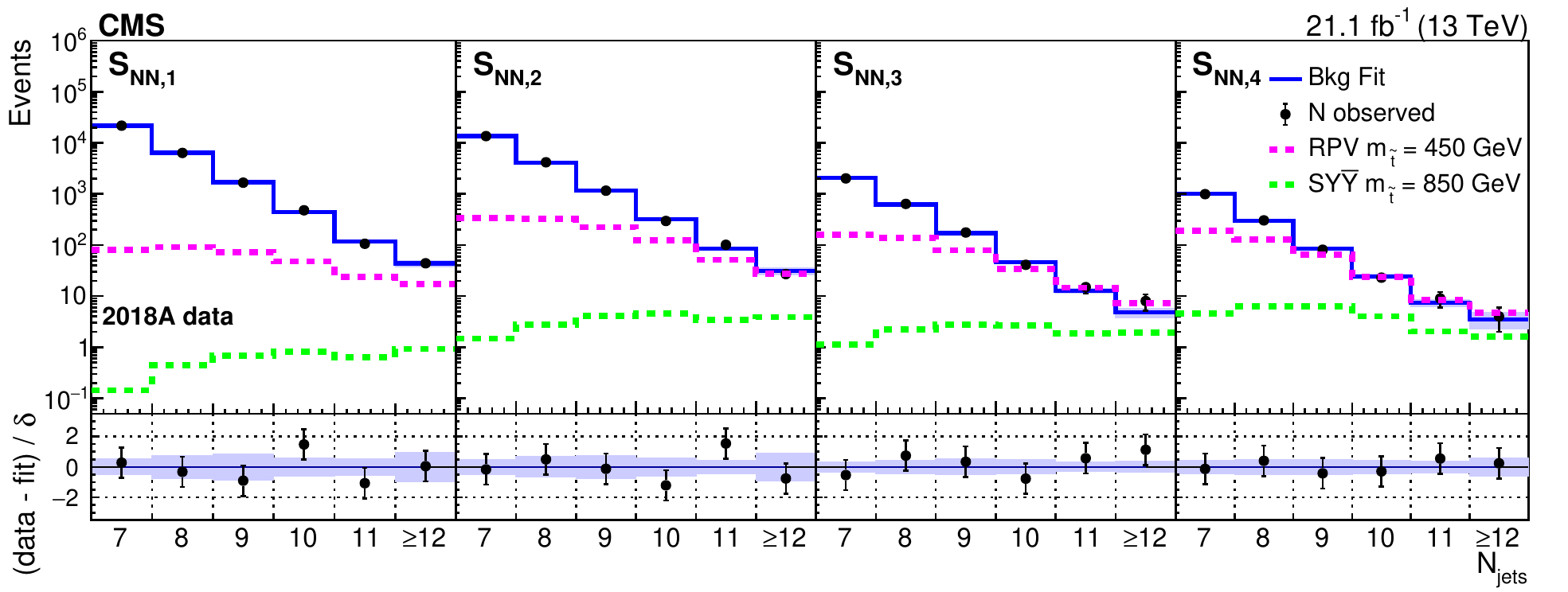}
\includegraphics[width=0.82\textwidth]{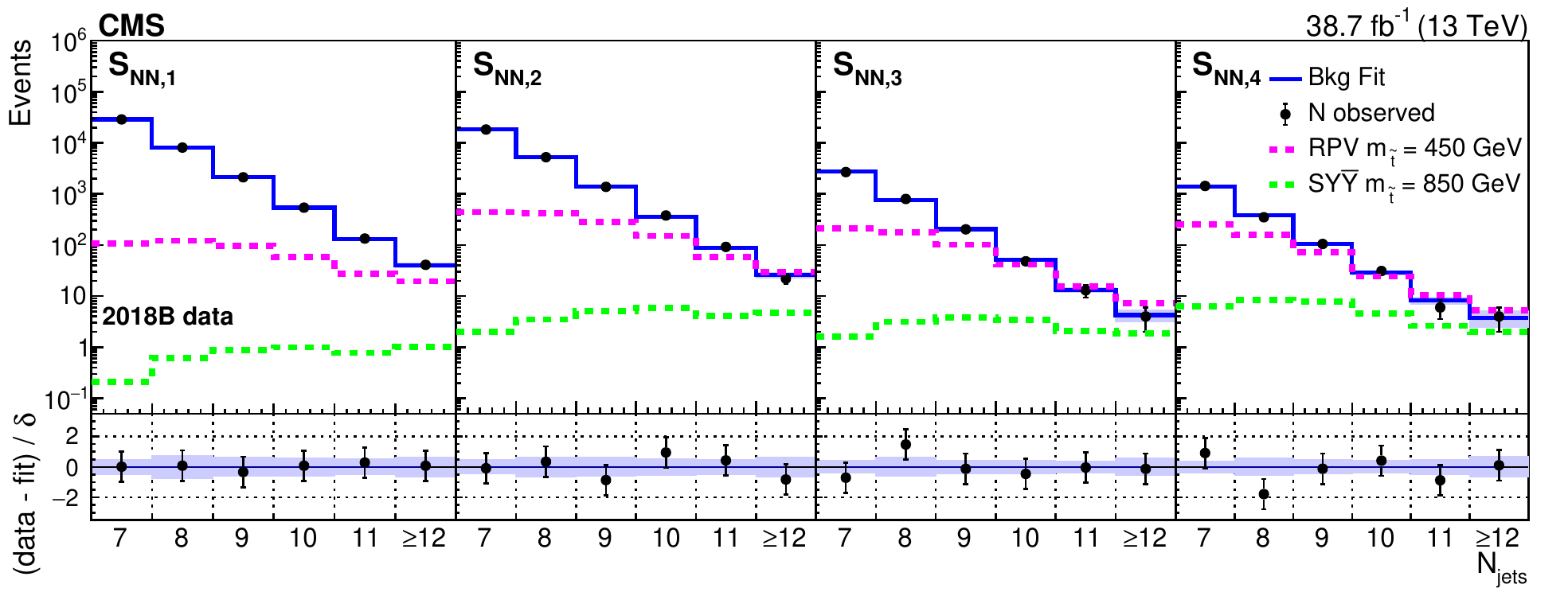}
\caption{
Fitted background prediction and observed data counts for 
2016, 2017, 2018A, and 2018B (from upper to lower rows) as functions of \njets in each 
of the four bins in \snn. The signal distributions normalized to 
the predicted cross section for the RPV model with $\mstop=450\GeV$ 
and the stealth \syy model with $\mstop =850\GeV$ are shown for comparison.
The lower panel of each plot displays the difference between the number of observed events and 
the number of events determined by the fit divided by the 
statistical uncertainty associated with the observed number of 
events ($\delta$) as black points with error bars 
denoting $\delta$.  The blue band shows the total systematic
uncertainty in the fit from all nuisance parameters.
\label{fig:results_fitPlots}}
\end{figure*}

\begin{figure}[tbp]
\centering
\includegraphics[width=0.49\textwidth]{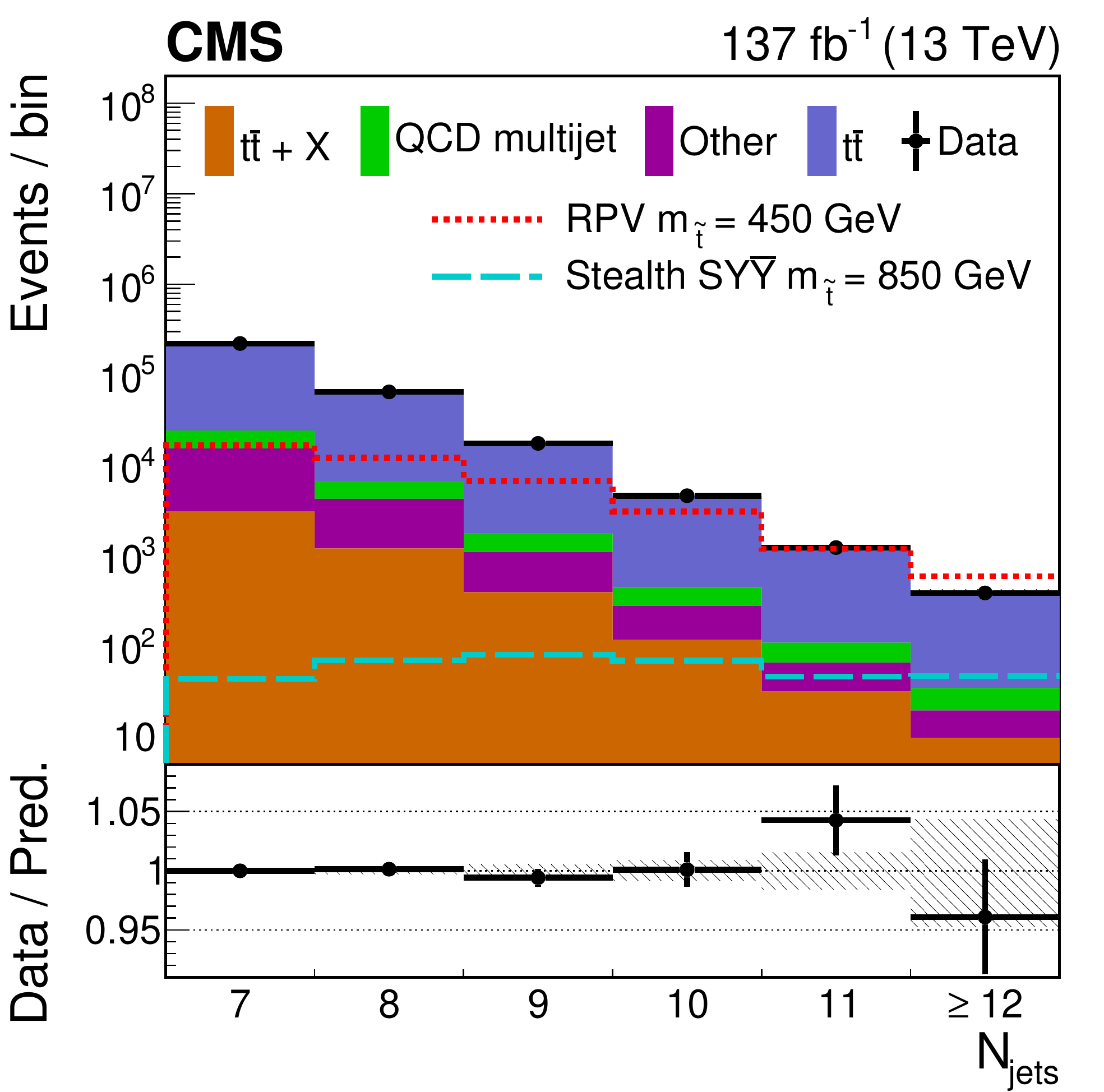}
\caption{
Background prediction from the background-only fit and 
observed data counts as a function of \njets summed 
over data periods and \snn bins. Overlaid are expected 
distributions for the RPV and stealth \syy models with 
$\mstop = 450$ and 850\GeV, respectively, normalized 
according to the top squark pair production cross section. 
For visualization purposes, the hatched band in the 
lower panel shows the quadrature sum of all of the 
uncertainties on the background prediction.
\label{fig:njets_datavsMC}}
\end{figure}

The results of the fit to 2016, 2017, 2018A, and 2018B 
data sets with the signal strength fixed to zero (background-only fit) 
are shown along with the observed number of events in 
Fig.~\ref{fig:results_fitPlots}; 
each column (row) in the plot array corresponds 
to a specific \snn bin (data set).
The expected distributions for top squark pair production
in the specific RPV ($\mstop = 450\GeV$) 
and stealth \syy models ($\mstop = 850\GeV$) 
described in Section~\ref{sec:intro} are 
overlaid for illustration purposes. 
The lower panel of each plot displays the difference between 
the observed number of events and the total 
number of expected events determined by the fit divided by the 
statistical uncertainty associated with the observed number of 
events ($\delta$) as black points with error bars 
denoting $\delta$.  The blue band shows the total 
uncertainty in the fit determined from the full fit 
covariance matrix in order to account for the
correlations among fit parameters.
Figure~\ref{fig:njets_datavsMC} shows the results of the 
same background-only fit summed over \snn bins and data 
periods with separate contributions from each background.

\begin{figure}[tbp]
\centering
\includegraphics[width=0.49\textwidth]{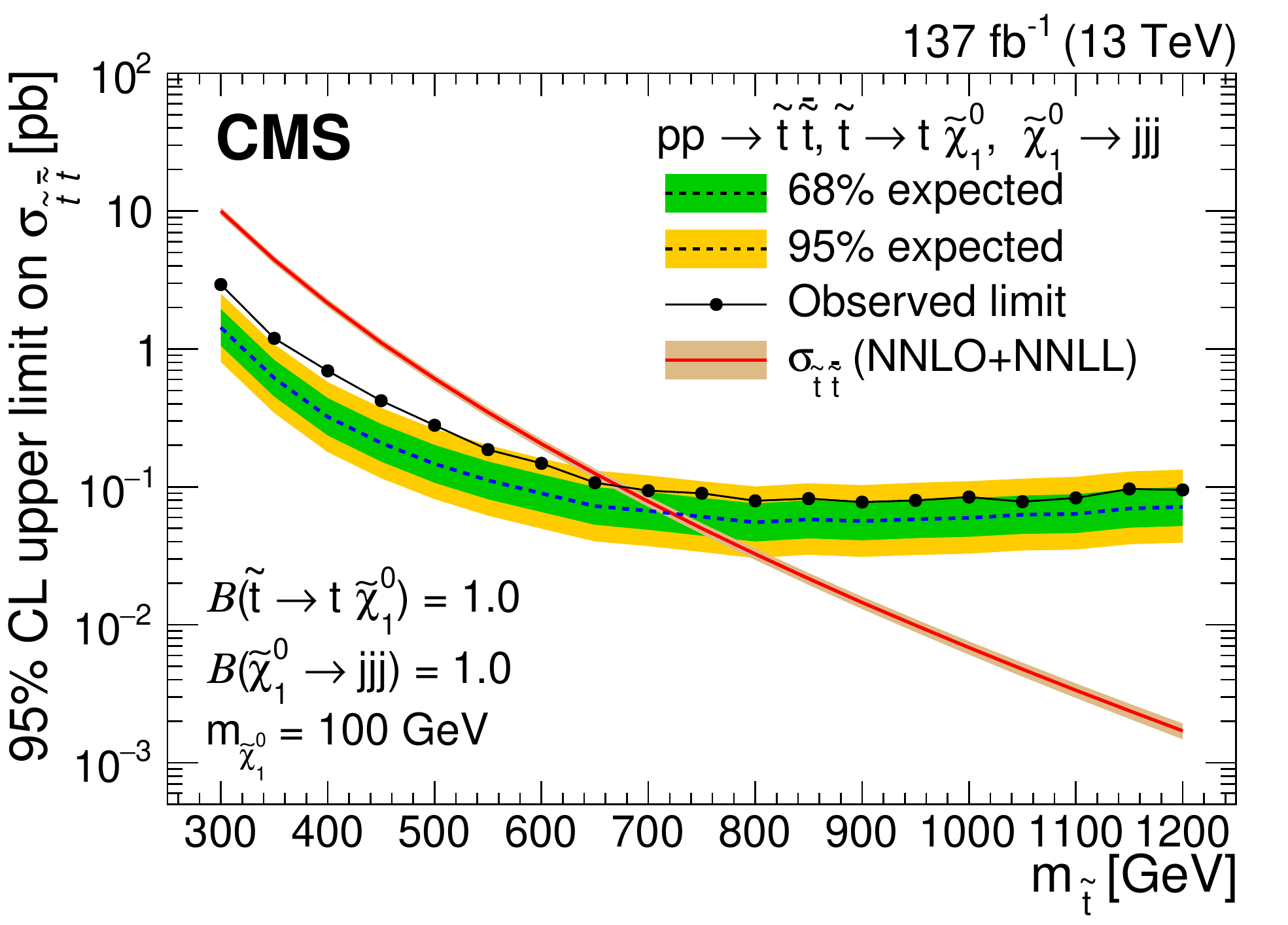}
\includegraphics[width=0.49\textwidth]{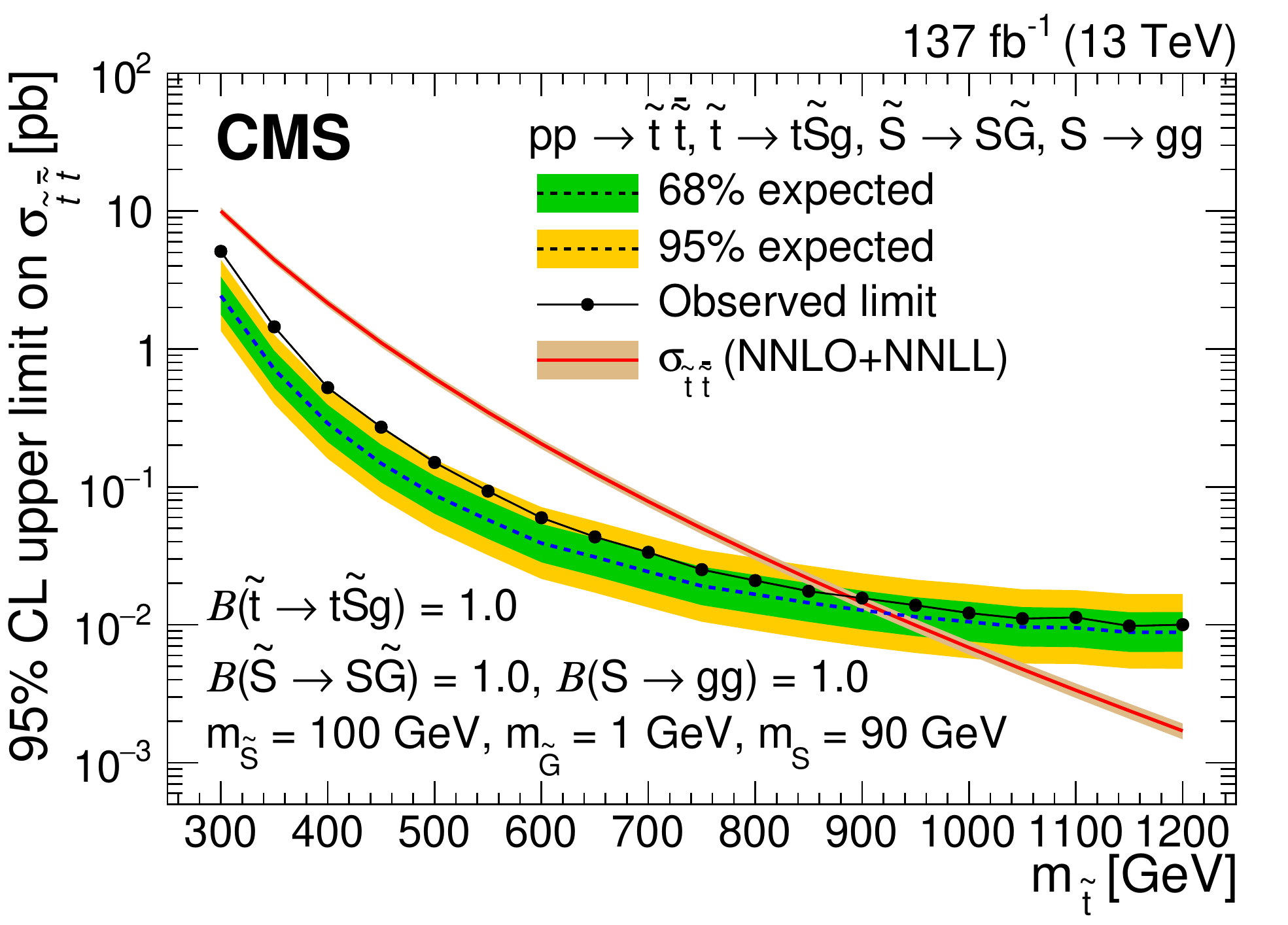}
\caption{ 
Expected and observed 95\% \CL upper limit on the top squark pair 
production cross section as a function of the top squark mass 
for the RPV (\cmsLeft) and stealth \syy (\cmsRight) SUSY models.
Particle masses and branching fractions assumed for each model 
are included on each plot.
The expected cross section computed at NNLO+NNLL accuracy 
is shown in the red curve.  
\label{fig:SYY-limits}}
\end{figure}

\begin{figure*}[!thp]
\centering
\includegraphics[width=0.49\textwidth]{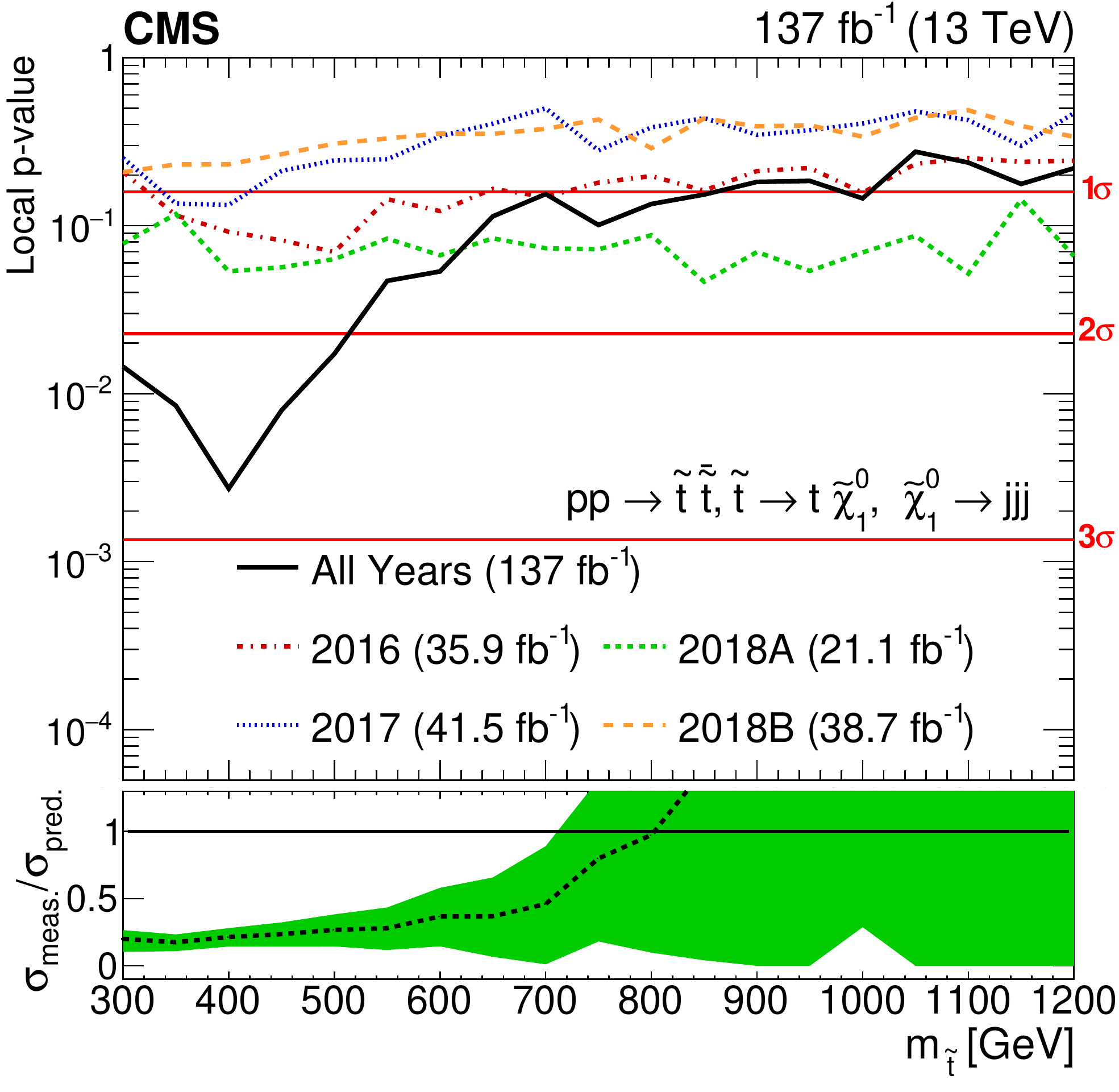}
\includegraphics[width=0.49\textwidth]{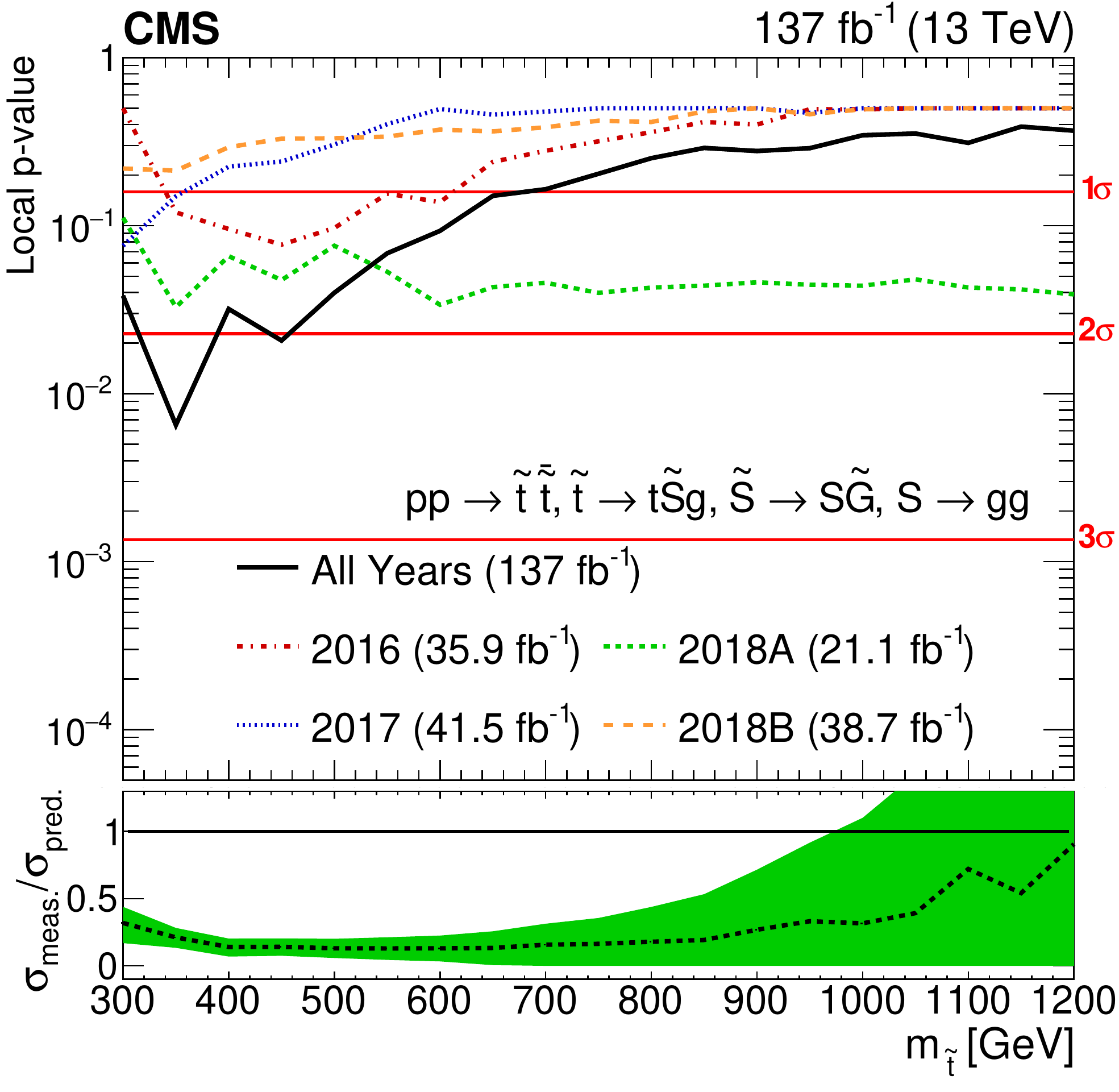}
\caption{Local $p$-value as a function of top squark mass for the 
RPV (\cmsLeft) and stealth \syy models (\cmsRight). The colored lines
show the $p$-values for separate fits of the 
2016 (red dash dotted), 2017 (blue dotted), 2018A (green short dashed), and 2018B (orange long dashed) data sets; 
the black line shows the $p$-value for the simultaneous fit of data sets.
The lower panels show the best fit signal strength ($\sigma_\text{meas.}/\sigma_\text{pred.}$) as a function of top 
squark mass with uncertainty denoted by the green band. 
\label{fig:pvaluePlots}}
\end{figure*}

The data are also used to determine the 95\% confidence 
level (\CL) upper limits on $\sigma_{\PSQt\PASQt}$ 
and the signal strength $p$-values~\cite{pvalue} 
for both the RPV and stealth \syy models obtained using the 
\CLs approach~\cite{CMS-NOTE-2011-005,CLS2,CLS1} 
with asymptotic formulae~\cite{Cowan:2010js} and 
the profile likelihood ratio as the test statistic. 
Figure~\ref{fig:SYY-limits}
shows the expected and observed cross section limits 
as a function of \mstop for the benchmark RPV and 
stealth \syy signal models.  
Comparing to the predicted cross section, these limits correspond
to the exclusion of top squark 
masses in the range 300--670 and 300--870\GeV for 
the benchmark RPV and stealth \syy models, respectively.
Figure~\ref{fig:pvaluePlots} shows the local 
$p$-value~\cite{pvalue} of the signal strength,
as a function of \mstop,
obtained from fits to the data with each signal 
strength as a free parameter for both the 
RPV and stealth \syy models. The $p$-value quantifies 
the probability for the background to produce an 
upward fluctuation at least as large as that observed.
Fits are performed and $p$-values obtained 
separately for each data set, as well as in 
a simultaneous fit to all data sets.
We observe the most extreme $p$-value to be 0.003, 
which corresponds to a local significance of 2.8\unit{$\sigma$}
and a best fit signal strength of $0.21 \pm 0.07$ for 
the RPV model with $\mstop = 400\GeV$ assuming unity 
branching fractions for the decays described in Section~\ref{sec:intro}.

\begin{figure*}[tbp]
\centering
\includegraphics[width=0.95\textwidth]{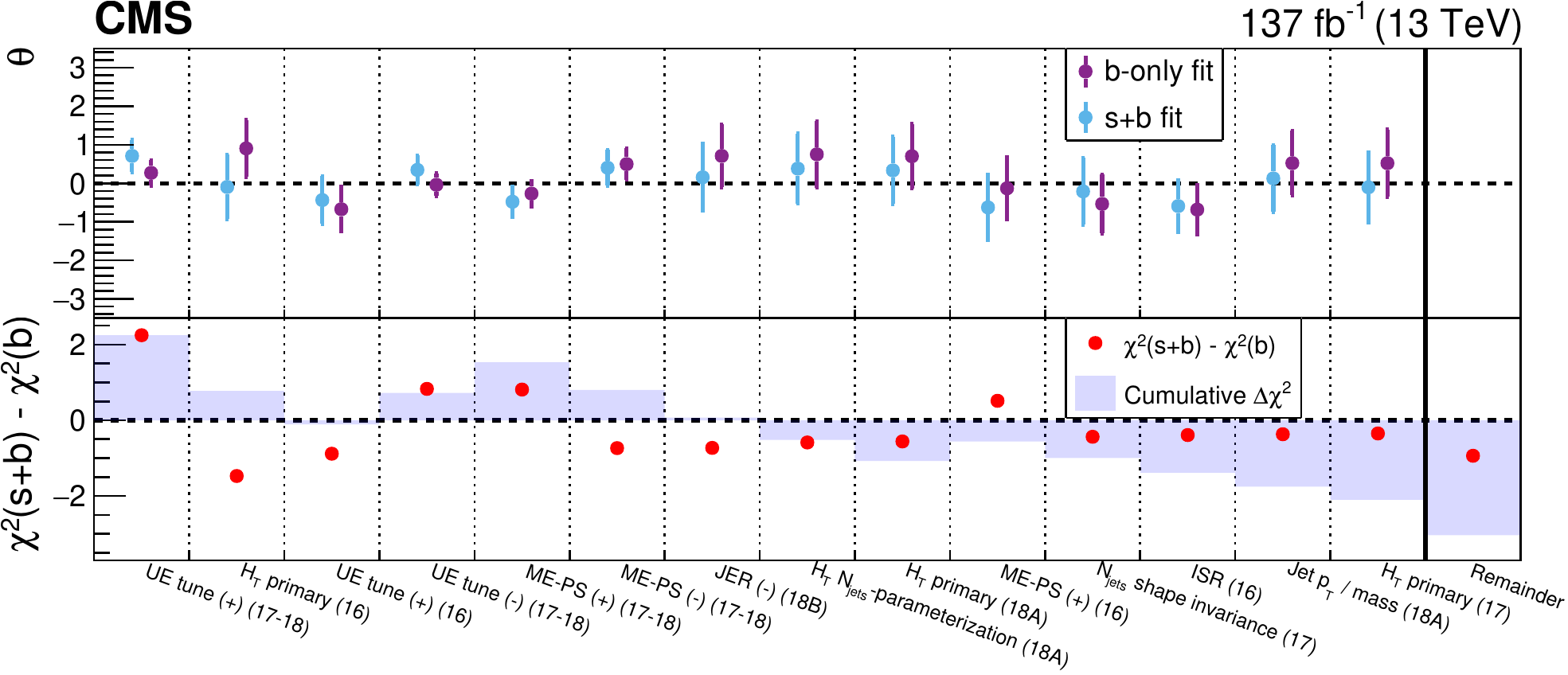}
\caption{
The upper panel shows the fit values ($\theta$) and 
uncertainties ($\delta_\theta$) for a selection 
of nuisance parameters (NP) from both the background-only fit (purple) and 
signal+background fit (blue) for the RPV model with $\mstop = 400\GeV$.  
The x-axis labels refer to the NP sources described in Section~\ref{sec:syst}, 
the data period (16, 17, etc.), and the direction of variation ($+$,$-$).
The lower panel shows the 
$\Delta\chi^{2}\equiv\chi^{2}(\mathrm{s}+\mathrm{b})-\chi^{2}(\mathrm{b})$ difference
of $\chi^2\equiv(\theta/\delta_\theta)^2$ 
from the signal+background (s+b) and background-only (b) fits 
as a red point for each NP and the cumulative sum
of $\Delta\chi^2$ from left to right as a blue shaded histogram.
The fourteen selected NP are those with $\vert\Delta\chi^2\vert>0.3$, 
and the NP are ordered from left to right by decreasing $\vert\Delta\chi^2\vert$.
The rightmost bin, separated by a vertical solid line, shows
the sum of $\Delta\chi^2$ for all NP not 
displayed in the figure (red point) and the sum of $\Delta\chi^2$
for all NP (blue shaded histogram).
\label{fig:nuisancePulls}}
\end{figure*}

The 2.8\unit{$\sigma$} local significance for the RPV model with $\mstop = 400\GeV$
is understood to arise from a combination of two effects.  First, although the level of 
agreement between the observed data and the background expectation shown 
in Fig.~\ref{fig:results_fitPlots} is reasonable, the agreement 
improves when the signal is included in the fit,
contributing approximately 1.1\unit{$\sigma$} to the significance. 
Second, the constrained nuisance parameters (NP) are pulled less 
from their initial values when the signal is included in the fit,
contributing approximately 1.7\unit{$\sigma$} to the significance.  
This second effect is illustrated in Fig.~\ref{fig:nuisancePulls} 
which shows for each of a selection of NP : the fit value 
($\theta$) and uncertainty ($\delta_\theta$) from both the background-only fit (b) 
and the signal+background fit (s+b), as well as the 
$\Delta\chi^2 \equiv \chi^2(\mathrm{s}+\mathrm{b}) - \chi^2 (\mathrm{b})$ difference 
of $\chi^2\equiv(\theta/\delta_\theta)^2$ from the two fits.  
A $\theta$ value of one indicates that the fit value of the nuisance parameter is one standard deviation from its nominal value, and a $\delta_\theta$ value less than one shows that the uncertainty is reduced in the fit relative to its initial value.
All NPs have $\theta$ values below one for the background-only fit, and several NPs related to \ttbar modeling are constrained with $\delta_\theta$ in the range of 0.25--0.40.
Figure~\ref{fig:nuisancePulls} also shows the cumulative and total sums of $\Delta\chi^2$ for the NPs, 
with the sum for all NP of $\sum\Delta\chi^2 = -3.0$ corresponding 
to an approximate contribution to the signal 
significance of $\sqrt{\vert\sum\Delta\chi^2\vert}=1.7\,\sigma$.

\section{Summary}
\label{sec:sum}

A first of its kind search for top squark pair production with subsequent decay 
characterized by two top quarks, additional gluons or light-flavor 
quarks, and low missing transverse momentum (\ptmiss) is described.  
Events containing exactly one electron or muon and at least seven jets, 
of which at least one should be \PQb tagged,
are selected from a sample of proton-proton collisions at $\sqrt{s}=13\TeV$ 
corresponding to an integrated luminosity of 137\fbinv 
collected with the CMS detector in 2016--2018. 
No requirement is made on \ptmiss. The dominant \ttbar background 
is predicted from data using a simultaneous fit of the 
jet multiplicity distribution across four bins of a neural 
network score. 

The results are interpreted in terms of top squark pair 
production in the context of $R$-parity violating (RPV) and 
stealth supersymmetry models.  
Top squark masses (\mstop) up to 670\GeV are excluded at 95\% confidence 
level for the RPV model in which 
the top squark decays to a top quark and the 
lightest neutralino, which subsequently decays 
to three light-flavor quarks via an off-shell 
squark through a trilinear coupling \Lampp.  
Top squark masses up to 870\GeV are excluded 
for the stealth supersymmetry model in which the top 
squark decays to a top quark, three gluons, and a gravitino 
via intermediate hidden sector particles.
The maximum observed local significance is 2.8 standard deviations 
corresponding to a best fit signal strength of 
$0.21 \pm 0.07$ for the RPV model with $\mstop = 400\GeV$.

\begin{acknowledgments}
  We congratulate our colleagues in the CERN accelerator departments for the excellent performance of the LHC and thank the technical and administrative staffs at CERN and at other CMS institutes for their contributions to the success of the CMS effort. In addition, we gratefully acknowledge the computing centers and personnel of the Worldwide LHC Computing Grid and other centers for delivering so effectively the computing infrastructure essential to our analyses. Finally, we acknowledge the enduring support for the construction and operation of the LHC, the CMS detector, and the supporting computing infrastructure provided by the following funding agencies: BMBWF and FWF (Austria); FNRS and FWO (Belgium); CNPq, CAPES, FAPERJ, FAPERGS, and FAPESP (Brazil); MES (Bulgaria); CERN; CAS, MoST, and NSFC (China); COLCIENCIAS (Colombia); MSES and CSF (Croatia); RIF (Cyprus); SENESCYT (Ecuador); MoER, ERC PUT and ERDF (Estonia); Academy of Finland, MEC, and HIP (Finland); CEA and CNRS/IN2P3 (France); BMBF, DFG, and HGF (Germany); GSRT (Greece); NKFIA (Hungary); DAE and DST (India); IPM (Iran); SFI (Ireland); INFN (Italy); MSIP and NRF (Republic of Korea); MES (Latvia); LAS (Lithuania); MOE and UM (Malaysia); BUAP, CINVESTAV, CONACYT, LNS, SEP, and UASLP-FAI (Mexico); MOS (Montenegro); MBIE (New Zealand); PAEC (Pakistan); MSHE and NSC (Poland); FCT (Portugal); JINR (Dubna); MON, RosAtom, RAS, RFBR, and NRC KI (Russia); MESTD (Serbia); SEIDI, CPAN, PCTI, and FEDER (Spain); MOSTR (Sri Lanka); Swiss Funding Agencies (Switzerland); MST (Taipei); ThEPCenter, IPST, STAR, and NSTDA (Thailand); TUBITAK and TAEK (Turkey); NASU (Ukraine); STFC (United Kingdom); DOE and NSF (USA).
  
  \hyphenation{Rachada-pisek} Individuals have received support from the Marie-Curie program and the European Research Council and Horizon 2020 Grant, contract Nos.\ 675440, 724704, 752730, and 765710 (European Union); the Leventis Foundation; the Alfred P.\ Sloan Foundation; the Alexander von Humboldt Foundation; the Belgian Federal Science Policy Office; the Fonds pour la Formation \`a la Recherche dans l'Industrie et dans l'Agriculture (FRIA-Belgium); the Agentschap voor Innovatie door Wetenschap en Technologie (IWT-Belgium); the F.R.S.-FNRS and FWO (Belgium) under the ``Excellence of Science -- EOS" -- be.h project n.\ 30820817; the Beijing Municipal Science \& Technology Commission, No. Z191100007219010; the Ministry of Education, Youth and Sports (MEYS) of the Czech Republic; the Deutsche Forschungsgemeinschaft (DFG), under Germany's Excellence Strategy -- EXC 2121 ``Quantum Universe" -- 390833306, and under project number 400140256 - GRK2497; the Lend\"ulet (``Momentum") Program and the J\'anos Bolyai Research Scholarship of the Hungarian Academy of Sciences, the New National Excellence Program \'UNKP, the NKFIA research grants 123842, 123959, 124845, 124850, 125105, 128713, 128786, and 129058 (Hungary); the Council of Science and Industrial Research, India; the HOMING PLUS program of the Foundation for Polish Science, cofinanced from European Union, Regional Development Fund, the Mobility Plus program of the Ministry of Science and Higher Education, the National Science Center (Poland), contracts Harmonia 2014/14/M/ST2/00428, Opus 2014/13/B/ST2/02543, 2014/15/B/ST2/03998, and 2015/19/B/ST2/02861, Sonata-bis 2012/07/E/ST2/01406; the National Priorities Research Program by Qatar National Research Fund; the Ministry of Science and Higher Education, project no. 0723-2020-0041 (Russia); the Programa Estatal de Fomento de la Investigaci{\'o}n Cient{\'i}fica y T{\'e}cnica de Excelencia Mar\'{\i}a de Maeztu, grant MDM-2015-0509 and the Programa Severo Ochoa del Principado de Asturias; the Thalis and Aristeia programs cofinanced by EU-ESF and the Greek NSRF; the Rachadapisek Sompot Fund for Postdoctoral Fellowship, Chulalongkorn University and the Chulalongkorn Academic into Its 2nd Century Project Advancement Project (Thailand); the Kavli Foundation; the Nvidia Corporation; the SuperMicro Corporation; the Welch Foundation, contract C-1845; and the Weston Havens Foundation (USA).
\end{acknowledgments}
\bibliography{auto_generated}

\cleardoublepage \appendix\section{The CMS Collaboration \label{app:collab}}\begin{sloppypar}\hyphenpenalty=5000\widowpenalty=500\clubpenalty=5000\input{SUS-19-004-authorlist.tex}\end{sloppypar}
\end{document}

%% file: SUS-19-004-authorlist.tex
\vskip\cmsinstskip
\textbf{Yerevan Physics Institute, Yerevan, Armenia}\\*[0pt]
A.M.~Sirunyan$^{\textrm{\dag}}$, A.~Tumasyan
\vskip\cmsinstskip
\textbf{Institut f\"{u}r Hochenergiephysik, Wien, Austria}\\*[0pt]
W.~Adam, J.W.~Andrejkovic, T.~Bergauer, S.~Chatterjee, M.~Dragicevic, A.~Escalante~Del~Valle, R.~Fr\"{u}hwirth\cmsAuthorMark{1}, M.~Jeitler\cmsAuthorMark{1}, N.~Krammer, L.~Lechner, D.~Liko, I.~Mikulec, F.M.~Pitters, J.~Schieck\cmsAuthorMark{1}, R.~Sch\"{o}fbeck, M.~Spanring, S.~Templ, W.~Waltenberger, C.-E.~Wulz\cmsAuthorMark{1}
% ...
\vskip\cmsinstskip
\dag: Deceased\\
1:  Also at Vienna University of Technology, Vienna, Austria\\
% ...
